\documentclass[usenatbib]{emulateapj}
\usepackage{subfigure}
\newcommand{\Msun}{M$_{\sun}$}

\shorttitle{Size-Magnitude Relation Evolution}
\shortauthors{Brooks et al.}

\begin{document}

\title{Interpreting the Evolution of the Size -- Luminosity Relation for Disk Galaxies from Redshift 1 to the Present}
\author{A.\,M. Brooks\altaffilmark{1,2},
        A.\,R. Solomon\altaffilmark{1,3},
        F. Governato\altaffilmark{4},
	J. McCleary\altaffilmark{5}
 	L.\,A. MacArthur\altaffilmark{6,7}
	C.\,B.\,A. Brook\altaffilmark{8}
        P. Jonsson\altaffilmark{9},
        T.\,R. Quinn\altaffilmark{4},
        J. Wadsley\altaffilmark{10},
}
\altaffiltext{1}{California Institute of Technology, M/C 350-17, Pasadena, CA, 91125}
\altaffiltext{2}{e-mail address: abrooks@tapir.caltech.edu }
\altaffiltext{3}{Astronomy Department, Yale University, P.O.\,Box 208101, New Haven, CT 06520}
\altaffiltext{4}{Astronomy Department, University of Washington, Box 351580, Seattle, WA, 98195}
\altaffiltext{5}{Department of Astronomy, New Mexico State University, P.O.\,Box 30001, MSC 4500, Las Cruces, New Mexico 88003}
\altaffiltext{6}{NRC Herzberg Institute of Astrophysics, Victoria, BC, V9E 2E7, Canada}
\altaffiltext{7}{Department of Physics \& Astronomy, University of Victoria, Victoria, BC, V8P 1A1, Canada}
\altaffiltext{8}{Jeremiah Horrocks Institute, University of Central Lancashire, Preston, Lancashire, PR1 2HE, United Kingdom}
\altaffiltext{9}{Harvard-Smithsonian Center for Astrophysics, 60 Garden Street, Cambridge, MA 02138}
\altaffiltext{10}{Department of Physics and Astronomy, McMaster University, Hamilton, Ontario, L88 4M1, Canada}


\begin{abstract}

A sample of very high resolution cosmological disk galaxy simulations 
is used to investigate the evolution of galaxy disk sizes back to 
redshift 1 within the $\Lambda$CDM cosmology.  Artificial images in 
the rest frame $B$ band are generated, allowing for a measurement of 
disk scale lengths using surface brightness profiles as observations 
would, and avoiding any assumption that light must follow mass as 
previous models have assumed.  We demonstrate that these simulated disks 
are an excellent match to the observed magnitude -- size relation for 
both local disks, and for disks at $z$=1 in the magnitude/mass range 
of overlap.  We disentangle the evolution seen in the population as a 
whole from the evolution of individual disk galaxies.  In agreement with 
observations, our simulated disks undergo roughly 1.5 
magnitudes/arcsec$^{2}$ of surface brightness dimming since $z$=1.  
We find evidence that evolution in the magnitude -- size plane varies 
by mass, such that galaxies with M$_*$ $\geq$ 10$^9$ \Msun ~undergo more 
evolution in size than luminosity, while dwarf galaxies tend to evolve 
potentially more in luminosity.  The disks grow in such a way as to stay 
on roughly the same stellar mass -- size relation with time.  Finally, 
due to an evolving stellar mass -- SFR relation, a galaxy at a given 
stellar mass (or size) at $z$=1 will reside in a more massive halo and 
have a higher SFR, and thus a higher luminosity, than a counterpart of 
the same stellar mass at $z$=0.

\end{abstract}



\setcounter{footnote}{0}

\section{Introduction}
\label{intro}

Recent observational surveys \citep[e.g., {\sc COSMOS, GOODS, GEMS, 
SINGS, SDSS},][]{cosmos, goods, gems, sings, sdss} have allowed us for 
the first time to statistically explore issues of galaxy formation. 
One of the immediate challenges to disk formation theory that is presented 
by these observations comes in the form of evidence that there has been 
little change in the sizes of disk galaxies since $z$=1, despite an 
expectation that disks should be growing in size over this time.  
Observations tell us that there exist large disks by $z$=1 (before the 
universe was even half of its present age), and suggesting that these 
disks must be assembled prior to this epoch \citep[e.g.,][]{vogt96, 
roche98, lilly98,simard99,rav04,ferguson04,trujillo04,barden05,sargent07,
melbourne07,kanwar08}.  Recent observations lend strong support to the 
existence of disks back to even higher redshifts \citep{labbe03, fs06, shapiro08, 
stark08, wright09, jones09}, and for disk assembly at even higher redshifts 
in the form of clump-chain galaxies in the Hubble Ultra Deep Field 
\citep{elmegreen05, elmegreen07}. 

Results from large surveys suggest that there has been only weak evolution 
in disk sizes since $z$=1.  There is no evidence that the 
size function of disks has evolved back to $z$=1 \citep{lilly98, rav04, 
kanwar08}.  \citet{kanwar08} found that the shape of the size function 
did not evolve with redshift, though the normalization (or amplitude) did.  
This can be interpreted in two ways.  First, the normalization will vary 
if the number density of disks varies with time.  However, observations 
suggest that the number density of disks is constant to $z$=1 \citep{lilly98, 
sargent07}. 
Second, if it is assumed that galaxies dim with time (e.g., due to a 
passively evolving stellar population or decline in star formation rate), 
a 1 to 1.5 magnitude dimming since $z$=1 could explain the change in 
normalization, while keeping the size distribution of disks constant 
\citep{kanwar08}.  Hence, there is no immediate evidence for a change 
in the sizes of galaxy disks with time.  

Studies of the evolution in the magnitude -- size relationship for disk 
galaxies back to $z$=1 have found similar results.  Importantly, these 
studies must use surface brightness evolution to interpret evolution 
within the magnitude -- size plane.  This is not straightforward, as an 
increase in disk scale lengths at a fixed magnitude between $z$=1 and $z$=0 
can be mimicked by a decrease in luminosity at a fixed size \citep{trujillo04}.  
Selection effects are difficult to disentangle, and studies that require high 
and low $z$ samples to adhere to the same selection biases have found that 
large disks are consistent with no evolution in surface brightness 
\citep{simard99, rav04}.  Later studies concluded that this work was too 
restrictive, and that a careful treatment of completeness as a function of 
redshift supports surface brightness dimming over time \citep{schade96, 
roche98, lilly98, bs02, rav04, trujillo05, barden05, melbourne07, kanwar08}.  
Most studies conclude that pure size evolution at a fixed magnitude is ruled 
out \citep{melbourne07}.  Instead, the observed surface brightness evolution 
is best explained by luminosity dimming, with the amount of dimming dependent 
on galaxy size so that lower mass galaxies have undergone more dimming 
since $z$=1 than massive galaxies \citep{melbourne07, kanwar08}.  However, 
some amount of size evolution can't be ruled out, and at least weak size 
evolution is favored \citep{reshetnikov03, barden05, trujillo05}. 

While magnitude is assumed to scale with the mass of a galaxy, it is 
obvious from the above discussion that magnitude at a given mass is not 
necessarily constant in time.  The more fundamental relation is between size 
and stellar mass.  Magnitude at a given stellar mass is likely to increase back 
to $z$=1 due to the fact that there is an evolving star formation rate (SFR) -- 
mass relation \citep{gavazzi96, boselli01, K03, brinchmann04, feulner05, Erb06, 
salim07, noeske07a, noeske07b, elbaz07, Daddi07, schiminovich07, cb08, pannella09, 
damen09a, damen09b, dunne09, rodighiero10, oliver10, mannucci10, laralopez10}. 
\citet{barden05} examined the stellar mass -- size relation for disk 
dominated galaxies back to $z$=1, 
and found weak or no evolution.  This result can be interpreted in two 
ways.  First, given the change of $\sim$1 magnitude of surface brightness 
dimming in the same population, the result is consistent with a passively 
evolving stellar population at a given mass, with no growth of galaxy disks. 
Second, it could imply that galaxies are growing, but that they must grow 
in such a way as to evolve {\it along} the stellar mass -- size relation 
with time.  

On the other hand, galaxy disk formation theory predicts that the sizes of 
disks, both individual and as a population, should be growing since $z$=1.
In the standard picture, gas in a halo conserves its specific angular 
momentum (equal to that of the dark matter) as it cools to form a 
centrifugally supported disk that grows from the inside out \citep{wr78, 
FE80}.  In the simplest model, in which the density profile of galaxies is 
modeled as a singular isothermal sphere (SIS), the radius of the resulting 
disk is proportional to the parent halo virial radius, which grows inversely 
with the Hubble parameter, $H(z)^{-1}$.  
For a concordance $\Lambda$CDM cosmology \citep{wmap5, wmap7}, this relation 
predicts that disks at $z$=1 should be nearly a factor of two smaller than 
their $z$=0 counterparts at the same circular velocity, $V_c$ \citep{mmw, 
mao98, vdb98}.  

\citet{somerville08} adopted more reasonable assumptions to update the simple 
theoretical model that suggests that galaxy disks should grow by nearly a 
factor of two since $z$=1, primarily by adopting NFW density profiles 
rather than SIS profiles.  Unlike the SIS model, the concentrations of NFW 
profiles were lower in the past.  Combined with disk stability arguments 
\citep{stability}, this model predicts a much weaker evolution in disk sizes 
(15\%-20\%) at a fixed stellar mass, consistent with the \citet{barden05} 
results.  These semi-analytic models (SAMs) do not follow 
the evolution of individual disk galaxies with time.  Rather, they examine an 
instantaneous population of disks.  Clearly, this is the type of information 
derived from observations, but it prevents an interpretation of the evolution in 
individual disk galaxies based on observations of the population as a whole.
\citet{far09} instead follow the evolution of individual disks in their SAM, 
and demonstrate that disks tend to grow along the stellar mass -- size 
relation, with only weak evolution, consistent with the results of 
\citet{barden05}.

The analytic models discussed above \citep[e.g.,][]{mmw, somerville08, far09} 
have until recently been the only available theoretical tool with which to 
investigate the evolution of galactic disk sizes \citep[though see][]{brook06}.  
Simulations, with their ability to capture complex gas processes in mergers and 
subsequent SF, should be an ideal tool that allow a better trace of the 
distribution of stellar light, while SAMs must assume that light follows mass.  
However, simulations of disk galaxy formation in a CDM context have historically 
produced unrealistic disks that are too compact (dense), too small overall, and 
rotating too fast at a given radius \citep[e.g.,][]{sn99,ns00,eke01,Abadi03,
governato04}.  This failure has been named the ``angular momentum catastrophe.''  
This catastrophe has been largely associated with the overcooling problem, in 
which baryons cool too quickly at early times and become very dense and 
concentrated at the center of halos before merging. These halos then experience 
dynamical friction in subsequent mergers, and the resulting disks show the 
classic signs of the angular momentum catastrophe \citep{nb91, nw94, katz94, 
maller02b}. Thus, feedback mechanisms at early times have historically been 
invoked to prevent overcooling with moderate success \citep{DS86, sommerlarsen99, 
thacker00b, thacker01, sommerlarsen03, Robertson04, okamoto05, donghia06, 
keres09, sales10} but additional artificial exchange of angular momentum can 
still occur between baryons and dark matter in SPH simulations with low numerical 
resolution \citep{governato04, naab07, G07, kaufmann07, mayer08, piontek09a, 
sales10}.  Hence, to avoid the angular momentum catastrophe and produce realistic 
disks, simulations must have both a physically motivated feedback prescription, 
and very high numerical (mass and force) resolution \citep{cecilia1, cecilia2, 
zavala08, ceverino09}.

The simulations used in this study are part of a suite of very high numerical 
resolution disk galaxy simulations that incorporate a star formation (SF) and 
supernova (SN) feedback scheme that has been shown to overcome many of the 
past problems of disk galaxy simulations to successfully match a number of 
observed properties of galaxies (as discussed further below, and shown by the 
results of this paper).  

In this paper, we use simulations of individual disk galaxies to follow their 
evolution in the magnitude -- size plane and the stellar mass -- size plane.  We 
attempt to mimic the observations of these galaxies in a manner similar to the 
observed population, generating mock surface brightness images to measure disk 
scale lengths, avoiding the assumption that light must follow mass.  We 
demonstrate that the simulated galaxies have properties 
similar to real galaxies.  In \S2 we describe our small sample of very high 
resolution simulated disk galaxies that span a representative range of masses, 
merger histories, and spin values.  We show that our limited sample is in excellent 
agreement with the population of galaxies surveyed at both $z$=0 (\S3) and $z$=1 
(\S4).  In \S5, we disentangle the evolution of individual galaxies from the 
evolution seen in the population as a whole.  We conclude in \S6.

\section{Simulations \& Analysis}

\subsection{The Simulations}

\begin{figure}
\plotone{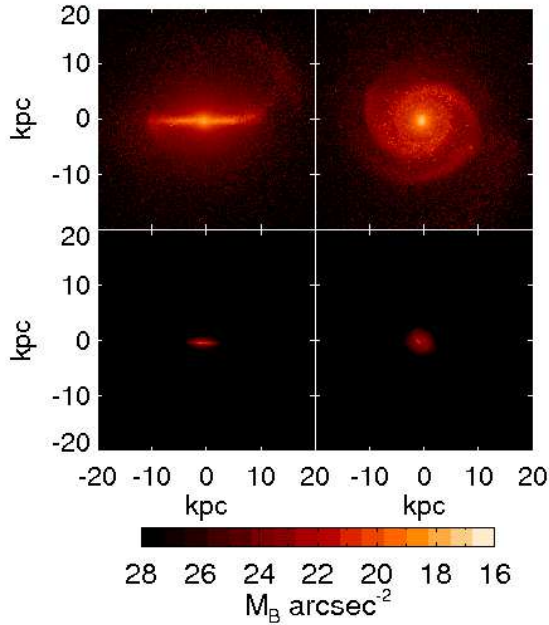}
\caption{Galaxies h277 (top) and h799 (bottom), seen edge on (left) 
and face on (right) at $z$=0 in the $B$ band.  These galaxies span the 
full range of virial masses presented in this paper, two orders of 
magnitude. }
\label{fig1}
\end{figure}

These simulations were run with the N-Body + Smoothed Particle 
Hydrodynamics (SPH) code {\sc Gasoline} \citep{pkdgrav, gasoline} in a 
fully cosmological $\Lambda$CDM context using WMAP year 3 
parameters\footnote{The choice of WMAP3 cosmology over WMAP7 cosmology 
has no impact on the results presented here.}
($\Omega_0=0.26$, $\Lambda$=0.74, $h=0.73$, $\sigma_8$=0.77, n=0.96).  
The galaxies were originally selected from either a 25Mpc or 50Mpc 
(depending on their mass) N-Body simulation with uniform mass resolution 
throughout, and then resimulated 
at higher resolution with baryons using the volume renormalization 
technique \citep{KW93, nw94}.  This technique allows for significantly 
higher resolution on the central galaxy while also capturing the effect 
of large scale torques that are thought to deliver angular momentum to 
the galaxy \citep{white84, barnes87}.  

The SF/SN scheme used in these simulations has been shown to effectively 
regulate star formation efficiency as a function of halo mass, resulting 
in a stellar mass -- metallicity relationship for the simulated galaxies 
that is in excellent agreement with observations both locally and at 
high $z$ \citep{Brooks07, maiolino08}.  This regulation of star formation 
also leads to realistic trends in gas fractions, with our lowest
mass galaxies being the most gas rich \citep{Brooks07}, 
reproducing the observed incidence rate of Damped Lyman $\alpha$ 
systems (QSO-DLAs) at $z$ = 3 \citep{pontzen08}, and the column 
densities of both QSO-DLAs and GRB-DLAs at $z$ = 3 \citep{pontzen08, 
pontzen10}.  Of critical importance for the present study is the fact that 
these simulated disks maintain sufficient angular momentum to match the 
observed Tully-Fisher relationship \citep{g08, merger} and produce galaxies 
with realistic disk sizes, as shown below. 

The full details of our physically motivated SN feedback implementation 
were originally presented in \citet{Stinson06}.  Briefly, the SF prescription 
ensures that the SFR density is a function of gas density according to 
the observed slope of the Kennicutt-Schmidt law, and a SF 
efficiency parameter, $c^*$, sets the normalization of this relation.  Each 
star particle represents a simple stellar population, born with a Kroupa 
initial mass function \citep{Kroupa}.  As massive stars go SN, energy and 
metals are deposited into the nearest neighbor gas particles.  The SN 
feedback recipe calculates the radius affected, and turns off cooling in 
those affected neighboring gas particles until the end of the snowplow 
phase as described by the Sedov-Taylor solution \citep{MO77}.  The amount 
of energy deposited amongst those neighbors is 0.4$\times$10$^{51}$ ergs, as 
was adopted in all of our previous work mentioned above.  Additionally, we 
include a uniform UV cosmic background following an updated model of 
\citet{HM96}.  

\citet{bulgeless} demonstrated that when force resolutions $\lesssim$100pc 
can be achieved, high density peaks that mimic SF regions in giant 
molecular clouds can be resolved with several hundreds of gas particles 
in $\sim$10$^6$ \Msun~clumps \citep[see also][]{booth07, ceverino09}.  This 
allows for the adoption of a realistic 
density threshold for SF \citep[100 amu/cm$^3$,][]{robertson08, tasker08, 
saitoh08}.  To match the Kennicutt-Schmidt law, the high density SF threshold 
must be offset by a slightly higher value of $c^*$ = 0.1.  This prescription 
leads to enhanced gas outflows that remove low angular momentum gas from the 
central regions of the galaxy \citep[][Brook et al., in prep.]{bulgeless}.  
At low masses, this creates a bulgeless disk with a linearly rising rotation 
curve, comparable to those observed \citep[Oh et al., in prep;][]{vdb01, 
thingsrot}.  The low 
mass galaxies used in this paper, drawn from a 25Mpc volume, can achieve 
force resolutions $\sim$100pc.  Hence, these simulations adopt the more 
realistic feedback prescription with high density SF and $c^*$ = 0.1.  
However, the more massive galaxies presented here are drawn from 50Mpc 
volumes, making it computational expensive to achieve similar force 
resolutions in a reasonable time.  Hence, these simulated galaxies cannot 
resolve the high density SF peaks easily, and a lower SF density threshold 
must be adopted, 0.1 amu/cm$^3$, with a lower $c^*$ = 0.05.  These latter 
values have been adopted in all of our previous work on MW mass galaxies 
\citep{Brooks07, G07, brooks09, merger, pontzen08, pontzen10}, and 
are a compromise that allow realistic disks to form, 
but are inefficient at driving gas outflows.  Thus, low angular 
momentum material that might be lost from the central regions is 
maintained, helping lead to the creation of large bulges in the MW 
mass galaxies \citep[see also][]{vdb01, bullock01, vdb02, maller02b, 
vdb03, donghia04, dutton09}.  Additionally, the creation of these large 
bulges can be due to missing physics.  In particular, AGN feedback has 
not yet been added to these simulations, and is potentially a key 
mechanism to help create realistic bulges.  Due to our small sample size, 
it is difficult to quantify if our bulges are substantially different from 
observed bulges in the same galaxy mass range.  

The simulated galaxies used in this work were selected to span a 
range of spin values and merger histories, with the last major merger 
redshift as low at 0.8.  Properties of these simulations are listed 
in Table~\ref{simsum}.

\begin{deluxetable*}{lcccccccccr}
\tablecaption{Simulated Galaxy Properties \label{simsum} }
\tablewidth{0pt}
\tablehead{\colhead{simulation} &
\colhead{M$_{vir}$} & \colhead {M$_{*}$} &
\colhead{M$^{DM}_{particle}$} &  \colhead{M$^{sph}_{particle}$} & \colhead{$\lambda$} &
\colhead{$\lambda_g$} & \colhead{z$_{lmm}$} &  \colhead{$\epsilon$} & \colhead{N within R$_{vir}$} \\
 & \colhead{\Msun} & \colhead {\Msun} & \colhead{\Msun} & \colhead{\Msun} &  &  &  & 
\colhead{pc} & \colhead{dm+star+gas} \\
\colhead{(1)} & \colhead{(2)} & \colhead{(3)} & \colhead{(4)} & \colhead{(5)} & 
\colhead{(6)} & \colhead{(7)} & \colhead{(8)} & \colhead{(9)} & \colhead{(10)} } 
\tabletypesize{\footnotesize}
\startdata
h516 & 3.9$\times$10$^{10}$ & 2.6$\times$10$^{8}$ & 1.6$\times$10$^{4}$ & 3.3$\times$10$^{3}$ & 0.05 & 0.05 & 1.2 & 87  & 3.5$\times$10$^{6}$ \\
h799 & 2.2$\times$10$^{10}$ & 1.3$\times$10$^{8}$ & 1.6$\times$10$^{4}$ & 3.3$\times$10$^{3}$ & 0.04 & 0.05 & 3.0 & 87  & 1.9$\times$10$^{6}$ \\
h603 & 3.8$\times$10$^{11}$ & 3.1$\times$10$^{10}$ & 3.0$\times$10$^{5}$ & 6.3$\times$10$^{4}$ & 0.06 & 0.11 & 1.1 & 231 & 3.8$\times$10$^{6}$ \\
h986 & 2.1$\times$10$^{11}$ & 2.2$\times$10$^{10}$ & 3.0$\times$10$^{5}$ & 6.3$\times$10$^{4}$ & 0.04 & 0.07 & 0.8 & 231 & 2.4$\times$10$^{6}$ \\
h239 & 8.5$\times$10$^{11}$ & 7.8$\times$10$^{10}$ & 1.2$\times$10$^{6}$ & 2.1$\times$10$^{5}$ & 0.03 & 0.05 & 1.1 & 347 & 2.8$\times$10$^{6}$\\
h258 & 8.0$\times$10$^{11}$ & 7.4$\times$10$^{10}$ & 1.2$\times$10$^{6}$ & 2.1$\times$10$^{5}$ & 0.04 & 0.07 & 0.8 & 347 & 2.8$\times$10$^{6}$\\
h277 & 7.1$\times$10$^{11}$ & 6.9$\times$10$^{10}$ & 1.2$\times$10$^{6}$ & 2.1$\times$10$^{5}$ & 0.03 & 0.04 & 3.0 & 347 & 2.3$\times$10$^{6}$ \\
h285 & 8.7$\times$10$^{11}$ & 8.1$\times$10$^{10}$ & 1.2$\times$10$^{6}$ & 2.1$\times$10$^{5}$ & 0.02 & 0.05 & 1.9 & 347 & 3.0$\times$10$^{6}$\\
\enddata
\tablecomments{Properties of the galaxies as drawn from the simulations.  
Columns (2) and (3) list the virial mass and total stellar mass of the halos at $z$=0.  
Columns (4) and (5) list the mass resolution of individual dark matter and star particles, 
respectively.  Column (6), $\lambda$, is the dimensionless spin parameter, $\grave{a}$ la 
\citet{bullock01}, for the entire halo.  Column (7) lists the spin parameter for all 
gas within the halo.    The last major merger (lmm) redshift in column (8) is defined at 
the time when the cores merge of two galaxies initially $\sim$3:1 in halo mass. 
Column (9), $\epsilon$, is the spline gravitational force softening. 
The final column (10) lists the total number of particles within 
the virial radius of the halo at $z$=0. } 
\end{deluxetable*}

\subsection{Deriving Disk Scale Lengths}
\label{scalelengths}

\begin{figure*}
\plotone{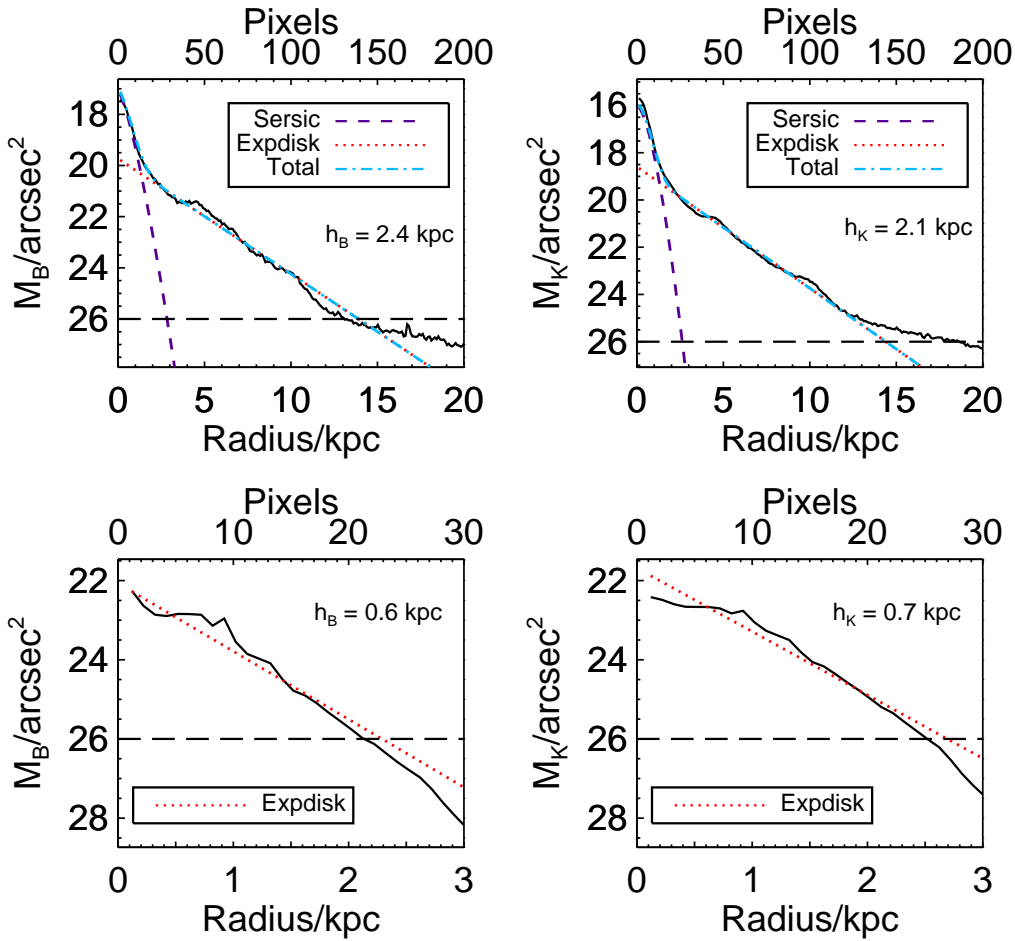}
\caption{Face-on, radially averaged $B$ band (left panels) and $K$ band 
(right panels) surface brightness profiles (solid lines) at $z$=0 for the 
same galaxies as in Fig.~\ref{fig1} (h277 and h799).  Bulge/disk 
decompositions resulting from {\sc Galfit} are shown, though the low 
mass galaxy in the bottom panel is bulgeless.  The long dashed line in 
each panel at 26 M$_B$/arcsec$^2$ represents the limiting surface 
brightness out to which the profiles were fit.}
\label{sb}
\end{figure*}

All of our simulations are of field galaxies that do not undergo major 
mergers below $z \sim$0.5, making them disk dominated at redshift zero.  
In order to compare our simulated disk scale lengths to observational 
results in as realistic a way as possible, we wish to fit the {\it 
light} profile rather than the underlying mass profile.  We create 
artificial surface brightness images of our simulated galaxies using 
{\sc Sunrise} \citep{sunrise, sunrise2}, a Monte Carlo radiative transfer 
program that produces a spectral energy distribution (SED) for each 
resolution element of the simulations.  This is done by identifying the 
age and metallicity of each star particle, which are then convolved with 
the Starburst99 stellar population synthesis models \citep{starburst99a, 
starburst99b} to produce an SED.  
{\sc Sunrise} assumes dust tracks with the metallicity of the gas 
particles, and performs ray tracing from each star particle to compute 
the observed SED including absorption and scattering.  From this,
we generate mock images in chosen filter bands.  Figure~\ref{fig1} 
shows artificial edge-on and face-on $B$ band images, including dust 
reprocessing, for two different mass galaxies.

Most of the observational surveys that have investigated the 
evolution of disk sizes worked in the rest frame $B$ band across 
redshifts.  To compare to these results as a function of redshift, we 
have generated face on $i$ band surface brightness images at $z$=0.5 and 
$z$=1 that include the effects of surface brightness dimming, and using 
rest frame face on $B$ band surface brightness images for the $z$=0 
galaxies.  Two of our disk galaxies are undergoing a major, 
disruptive merger at $z$=1.  For these galaxies, we created face on 
images for the most massive progenitor at a time just prior to the 
merger ($z$=1.25) while the disk was still rather undisturbed.  

The face on {\sc Sunrise} images were fit with a Sersic bulge component 
and an exponential disk component using the publicly available 2D surface 
brightness fitting code {\sc Galfit} \citep{galfit}.  Additionally, bars 
exist in both h603 and h986 at $z$=1 that were fit by a third component, 
reducing the B/D ratio of these galaxies at this step. 
The {\sc Galfit} results were checked against a 1D, radially averaged fit 
generated from the same images, and found to be in good agreement.  To 
mimic observations, the fits were required to be a good match down to a 
limiting surface brightness value of $\sim$26 mag/arcsec$^2$ (hence the 
importance of using surface brightness dimmed images at $z$=1).  Example 
fits are shown in Fig.~\ref{sb} for the same galaxies shown in 
Fig.~\ref{fig1}, with $K$ band fits shown simply for a comparison of the 
$B$ band results to those at a longer wavelength.  We 
verified that using redshifted, surface brightness dimmed {\sc Sunrise} 
images in the $i$ band at higher z yielded similar disk scale lengths as 
the rest frame $B$ band. 

As can be seen in Table~\ref{simsum}, a couple of these galaxies have 
undergone major mergers since $z$=1, while some galaxies have had a 
very quiescent history.  We note that at $z$=1, two of our galaxies 
(h239 and h285) had no component that could be well fit by an 
exponential.  This may be due to mergers that occur in these galaxies 
near this time.  Galaxy h239 has the most active merger history 
of our simulated galaxies, with continual mergers (both major and minor) 
until $z \sim$0.5.  Galaxy h285 begins to accrete a satellite with 
1/7 of the stellar mass of the main halo at $z$=1.25.  However, the core 
of this satellite does not merge with the main halo until $z \sim$0.8.  
In both cases, we searched in 250 Myr periods from 0.75 $< z <$ 1.25 to 
identify a step with a possible exponential disk, but none could be 
found.  Hence, those steps have been excluded from the high redshift 
analysis below.  

\begin{figure}
\plotone{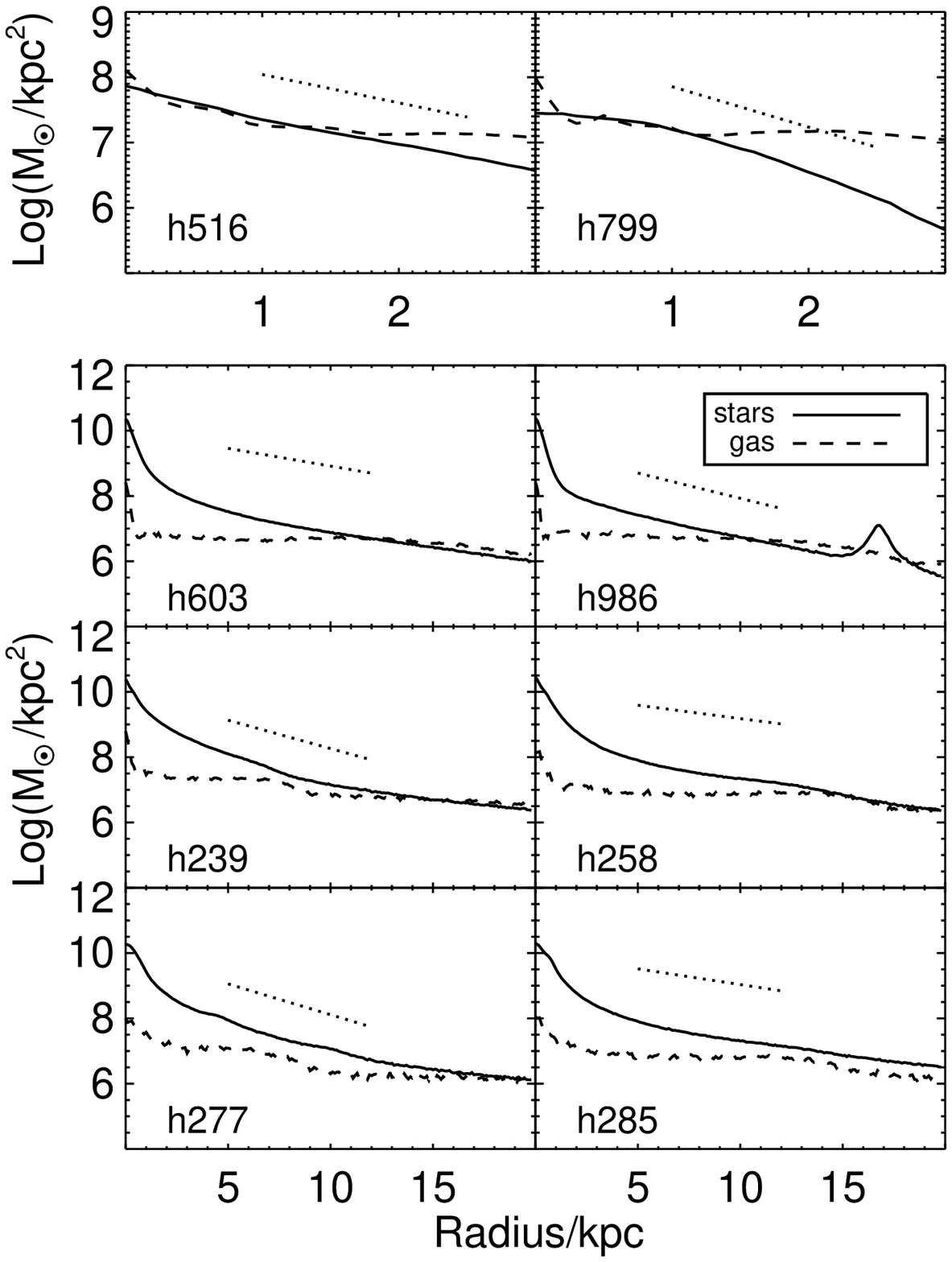}
\caption{The $z$=0 surface mass density distribution for the simulated 
galaxies.  Stellar surface density is shown by the solid lines, while gas 
surface density is shown by the dashed lines.  For comparison, the dotted 
lines in each panel show the $i$ band scale length derived using 
{\sc Sunrise/Galfit} for each galaxy (listed in Table~\ref{obssum}). } 
\label{ml}
\end{figure}

\citet{scannapieco10} demonstrated that measuring disk scale lengths 
based on light rather than a kinematic decomposition could dramatically 
increase the resulting disk-to-total, D/T, ratio of their simulated 
galaxies, making measurement technique a potentially important reconciler 
between observations and simulations.  In Fig.~\ref{ml} we plot the 
$z$=0 surface mass density as a function of radius for both the stars 
(solid line) and the gas (dashed line) in the simulated galaxies.  Also 
shown as the dotted line is the exponential fit derived in the $i$ band 
for these simulations (listed in Table~\ref{simsum}).  In a future paper 
(McCleary et al., in prep), we will explore in detail the differences 
between mass/kinematic results and light results (e.g., scale lengths, 
bulge-to-disk ratios, the role of dust and inclination, etc.).  Here, 
however, we note that an initial comparison of mass versus light 
exponential scale lengths demonstrates a less dramatic difference in 
our simulations than in \citet{scannapieco10}.  As seen in Fig.~\ref{ml}, 
the $i$ band scale length is generally a good match to the stellar surface 
mass density.  For a quantitative comparison, scale lengths were derived 
for the stellar mass over the same region that the disk dominates the 
light profile.  The two methods were found to agree to within 10\% for 
75\% of the galaxies.  
Variation in the techniques appears to be mostly attributable to 
varying mass-to-light ratios (McCleary et al., in prep).  

The bulge/disk decompositions were performed on the dust extinguished 
{\sc Sunrise} results.   However, disk scale lengths and central surface 
brightnesses were found to be identical in both the face on {\sc Sunrise} 
images with and without dust reprocessing, while the central bulge component 
was dust extinguished.  Thus, using the dust extinguished versus non dust 
extinguished decomposition has no effect on the results for the disk analysis 
presented below.  It is generally expected that dust should have some 
effect on both the central surface brightness and the scale length of the 
disk component, even for face on orientations \citep{calzetti01, graham08}.  
We speculate that our more massive galaxies have little/no effect from dust 
due to the fact that they are also more gas poor than observed galaxies 
(because they use up gas forming too many stars, discussed in more detail 
in Section~\ref{comp}).  Our low mass galaxies are gas rich, in agreement 
with observed galaxies in this mass range \citep{bulgeless, Geha06}, but 
metal poor \citep{Brooks07}, making dust negligible.  

\begin{figure*}
\plotone{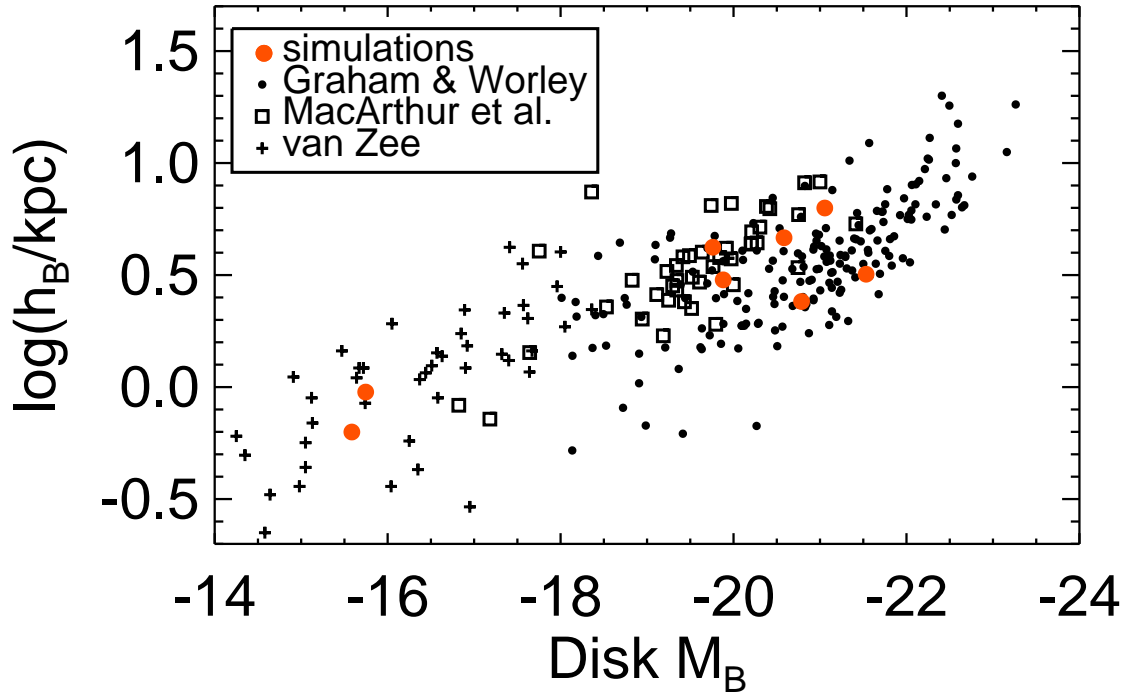} \caption{$B$ band disk scale length as a function 
of magnitude for our simulated galaxies.  Simulated galaxies at $z$=0 
are shown as large red circles.  The observational results of 
\citet{graham08}, \citet{macarthur03}, and \citet{vanzee00} are shown for 
comparison. }
\label{fig:z0}
\end{figure*}

It is important to note that the following analysis purposely focuses on 
the evolution of the {\it disk} alone.  That is, the discussion below of how 
properties evolve with time neglects the bulge component of the galaxy.  
This is done for three reasons.  First, as discussed above, our MW-mass 
galaxies tend to have larger bulges than observed.  This is a problem that 
has historically plagued MW-mass simulated disks \citep[e.g.,][]{Abadi03, 
cecilia2}, and is likely due to the inability to correctly model SF and 
feedback at the necessary high resolutions to drive loss of low angular 
momentum gas, and due to missing physics such as AGN feedback.  Hence, 
while the disks of our simulated galaxies appear to 
be in good agreement with observed disks (as presented below), the growth 
of the bulge in these massive galaxies is probably not modeled correctly.  
Second, observations of the evolution of disk galaxies in the 
magnitude -- size plane have either a) selected galaxies that can be fit with 
a small pure Sersic index, $n$, so that they are disk dominated, or b) 
done a bulge/disk decomposition.  In either case, the results we wish to 
compare to are concerned with disks only.  Finally, 
our goal here is to examine the growth of {\it disks} using these 
simulations, not bulges or spheroidal components.  

``Observable'' properties of the simulated galaxies are listed in 
Table~\ref{obssum}.  Total magnitudes in each band are derived from 
{\sc Sunrise}, with each galaxy being integrated out to roughly 50 comoving 
kpc (i.e., essentially all of the flux from the simulated galaxy is 
contained).  
Table~\ref{obssum} also lists the $z$=0 stellar mass derived for 
the entire galaxy (disk + spheroid) based on {\sc Sunrise} colors using the 
k\_sdss\_bell routine in the {\sc kcorrect} package \citep{kcorrect}.  A 
comparison to Table~\ref{simsum} shows that the simulated stellar masses 
are generally $\sim$40\% larger than those derived from the photometric 
results.  A more detailed comparison will be presented in McCleary et al., 
(in prep).
In Table~\ref{obssum} we list the bulge-to-disk ratios (B/D) 
despite the caveat mentioned above regarding large bulges in our higher 
mass galaxies.  Disk scale lengths in the $i$ band at $z$=0 are given in 
Table~\ref{obssum}, while the $B$ band scale lengths used in the 
following plots are given in Table~\ref{evol}.  As seen in observational 
studies \citep{dejong96a, macarthur03}, the $B$ band scale lengths tend to 
be longer than the $i$ band scale lengths, at least for the six most massive 
galaxies (with the two dwarf galaxies exhibiting different behavior).  It 
has been debated whether this trend for disk scale lengths to be longer at 
shorter wavelengths is due to dust effects or age gradients \citep{dejong96b}.  
As noted above, dust plays no role in our face on {\it disk} fits, meaning 
that this trend in the simulations is due entirely to age and metallicity 
gradients. 

\begin{deluxetable*}{lcccccccc}
\tablecaption{``Observable'' Galaxy Properties \label{obssum} }
\tablewidth{0pt}
\tablehead{\colhead{simulation} &
\colhead{M$_{i}$} & \colhead{M$_{B}$} & \colhead{B/D$_{i}$} & 
\colhead{B/D$_{B}$} & \colhead{$g-r$} & \colhead{M$^*_{kcorr}$} & 
\colhead{h$_{i}$} & \colhead{V$_{2.2}$} \\ 
   &  &  &  &  &  & \colhead{\Msun} & kpc & \colhead{km/s} \\
\colhead{(1)} & \colhead{(2)} & \colhead{(3)} & \colhead{(4)} & 
\colhead{(5)} & \colhead{(6)} & \colhead{(7)} & \colhead{(8)} & 
\colhead{(9)} } 
\tabletypesize{\footnotesize}
\startdata
h516 & -16.8 & -15.9 & 0.08 & 0.14 & 0.53 & 3.4$\times$10$^{8}$  & 1.0 & 51  \\
h799 & -16.2 & -15.6 & 0.00 & 0.00 & 0.37 & 1.1$\times$10$^{8}$  & 0.7 & 43  \\
h603 & -21.3 & -20.5 & 1.25 & 0.72 & 0.52 & 2.2$\times$10$^{10}$ & 4.0 & 143 \\
h986 & -21.1 & -20.3 & 0.63 & 0.38 & 0.45 & 1.3$\times$10$^{10}$ & 2.8 & 137 \\
h239 & -22.5 & -21.8 & 0.35 & 0.16 & 0.41 & 4.2$\times$10$^{10}$ & 2.5 & 246 \\
h258 & -22.4 & -21.5 & 0.86 & 0.45 & 0.50 & 5.3$\times$10$^{10}$ & 5.3 & 204 \\ 
h277 & -22.1 & -21.2 & 0.63 & 0.42 & 0.52 & 4.3$\times$10$^{10}$ & 2.3 & 250 \\
h285 & -22.3 & -21.3 & 1.33 & 0.78 & 0.55 & 5.6$\times$10$^{10}$ & 4.5 & 203 \\
\enddata
\tablecomments{Magnitude and color results are dust free measurements for the 
entire galaxy (disk and spheroidal components) at $z$=0.  B/D ratios include 
the effect of dust extinction (see discussion in Section~\ref{scalelengths}).  
The stellar masses listed in column (7) are 
derived using k\_sdss\_bell in the kcorrect 
package \citep{kcorrect} from SDSS $ugriz$ colors provided by {\sc 
Sunrise}.  Rotation curve velocities are measured at 2.2 $i$ band disk 
scale lengths. } 
\end{deluxetable*}

\section{The Fundamental Plane for Disk Galaxies}
\label{z0}

Size, magnitude (or mass), and rotational velocity make up a fundamental 
plane for disk galaxies \citep{pizagno05, gnedin07, courteau07}.  
\citet[][see their figure 5]{merger} demonstrated that these simulated 
galaxies lie on the observed magnitude -- velocity relation for disk 
galaxies (also known as the Tully-Fisher relation).  In this paper we 
derive disk scale lengths in order to extend this analysis, and show that 
our galaxies are also a good match to the observed magnitude -- size and 
velocity -- size relations as well.  All of our galaxies have sizes and 
velocities (Table~\ref{obssum}) that agree well with the observed velocity -- 
size trend \citep{courteau07}.  Two of the more massive galaxies have 
slightly high velocities for their size (h239 and h277), but still lie 
within the 2$\sigma$ scatter of observed galaxies. 

The $B$ band disk magnitudes, M$_B$, and scale lengths, h$_B$, of our 
simulated galaxies at $z$=0 are shown in Fig.~\ref{fig:z0}.  The simulated 
galaxies (red circles) are compared to three observational samples, 
\citet{macarthur03} and \citet{graham08} at the massive end, and 
\citet{vanzee00} at the low mass end.  The \citet{graham08} data have adopted 
a dust correction based on \citet{driver08}, while the \citet{macarthur03} 
data are dust corrected based on inclination, following equations 1, 2, and 7 
listed in \citet{graham08}.  Again, possibly due to low gas (and therefore 
dust) content, the six more massive galaxies shown here are free of dust 
effects.  As such, they should be compared to dust corrected observational 
data.  At the low mass end, however, 
\citet{vanzee00} makes no internal dust corrections, as the low metallicities 
of the dwarf galaxies in her sample are expected to lead to little effect from 
dust.\footnote{Note that there appears to be a slight offset between the dwarf 
galaxy sample and the higher mass sample.  This break has been observed 
previously, and suggests a structural difference between dwarf and larger 
disk galaxies \citep{schombert06}.} 
Clearly, the disk sizes of the simulated galaxies are in good agreement 
with observed disk sizes. 

We use the $B$ band scale lengths of the observational samples in 
Fig.~\ref{fig:z0}, which can be compared to the $i$ band results at $z$=1 
in the next section since the $B$ band is redshifted into the $i$ band at 
$z$=1.  However, the magnitude -- size (and stellar mass -- size) relation 
has been measured in multiple bands for a large sample of SDSS galaxies 
\citep{shen03, fathi10}.  The SDSS results are consistent with the results 
plotted in Fig.~\ref{fig:z0}, though we purposely select observational 
samples that have done a bulge/disk decomposition and fit a pure  
exponential to the disk surface brightness fit, in order to isolate the  
evolution of the disk and eliminate any contamination from a central 
spheroid.  

As discussed in the Introduction, historically it has been a challenge 
for cosmological disk galaxy simulations to match the observed fundamental 
relations for galaxy disks \citep[e.g.,][]{ns00}.  The failure to reproduce 
observed trends is the result of the angular momentum catastrophe,  
exacerbated by the fact that previous studies followed mass/kinematic 
results rather than light profiles.  

Our success at matching the observed disk scaling relations is a result 
of the fact that our disk baryons maintain angular momentum through time, 
due to an increased resolution and a physically motivated feedback 
mechanism \citep[see][for a review of how each of these processes 
contribute to the formation of realistic disks]{bashfest}.  A number of 
works have attempted to isolate the separate role of resolution 
and feedback \citep[e.g.,][]{governato04, kaufmann07, naab07, piontek09a}.  
In N-Body + SPH simulations, dark matter particles are typically an order 
of magnitude more massive than the gas particles, leading to an exchange 
of kinetic energy in two body interactions that kinematically heats the 
disk, randomizing velocity vectors, and potentially turning a disk 
component into a spheroid \citep{sw97, mayer08}. This effect can be 
reduced at higher mass resolutions that lower the discrepancy in 
particle masses. \citet{kaufmann07} used controlled (non-cosmological) 
N-Body+SPH simulations to evaluate the amount of angular momentum that 
can be lost purely due to resolution.  They concluded that 10$^6$ particles 
within the virial radius are necessary for a disk galaxy to maintain 
roughly 90\% of its original angular momentum.

Yet even if resolution can be increased to the point of maximum 
angular momentum retention, disks will still suffer dramatic angular 
momentum loss compared to observed disk galaxies if energy feedback is 
neglected.  Feedback prevents rapid, early cooling of gas particles 
(the overcooling problem discussed in the Introduction; see references 
therein).  Heating and expansion of the gas creates a hot reservoir, 
allowing the gas to cool at later times after the era of rapid mergers, 
and preventing angular momentum loss via dynamical friction in mergers 
\citep[e.g.,][]{maller02b}.  Additionally, by preventing overcooling, 
feedback prevents gas from rapidly turning into collisionless star 
particles.  Without feedback, halos of all masses are equally efficient 
at converting gas into stars \citep{Brooks07}, producing galaxies that 
are too gas poor compared to $z$ = 0 disk galaxies.  Because feedback 
regulates star formation, it allows for gas reservoirs to develop that 
allow disks to survive to the present day \citep{hopkinsbulgea, merger, 
moster10b}.  The existence of large, thin disks at the present day 
thus requires feedback.

Not all SN feedback schemes lead to disk galaxies that satisfy observed 
constraints such as the Tully-Fisher relation, size -- velocity or 
size -- luminosity relations, or the stellar mass -- metallicity relation.  
Because the nearest neighbor gas particles surrounding SNe are dense and cold,
a simple energy deposition will quickly be radiated away and not affect 
the simulation \citep[e.g.,][]{katz92, sn99}.  Two main schemes have 
been adopted to overcome this problem.  In the first, a multiphase 
model of the ISM is implemented with a hot gas reservoir inside each 
gas particle \citep{hultman99, marri03, springel03, cecelia0,harfst06}, 
overcoming the problem of thermal energy being spread over the entire 
particle mass, and preventing the hot gas particles from being 
artificially influenced by their cold gas nearest neighbors.
In the second, cooling is turned off in the gas particles near a SN 
explosion in order to mimic the sub-resolution adiabatic expansion of 
the SN \citep{thacker00b, thacker01}.  The ``blastwave'' scheme 
adopted here also turns off cooling in nearest neighbor particles, but
attempts to model this based on what is known about actual SNe,
determining the radius of each SN remnant based on the analytic
blastwave solution for a SN remnant \citep{MO77, Stinson06}, and
cooling is only turned off for those particles within the blast
radius.  Because many SNe typically contribute feedback within a
dense star forming region, the thermal energy from all of these SNe
can combine to create a larger blast radius.  The differences in the 
resulting disk between a simple energy deposition (``thermal'' 
feedback) and the blastwave model have been examined in 
\citet{mayer08} and \citet{g08}.  As examined by these previous 
works, the adoption of the ``blastwave'' feedback model, combined 
with high resolution (all simulations presented here have more than 
10$^6$ particles within the virial radius at $z$ = 0, see 
Table~\ref{simsum}), overcome past problems with cosmological disk 
galaxy simulations and allow for the present study.

\section{Evolution of the Magnitude -- Size Relation}
\label{magsize}

Having established that these simulated disk galaxies match observed 
disk scaling relations at $z$=0, we now use them to investigate 
the degeneracy in the magnitude -- size evolution with time.  First, 
we establish that the simulations also match the available observational 
data at $z$=1.

\subsection{Evolution as a Population}

\begin{figure}
\plotone{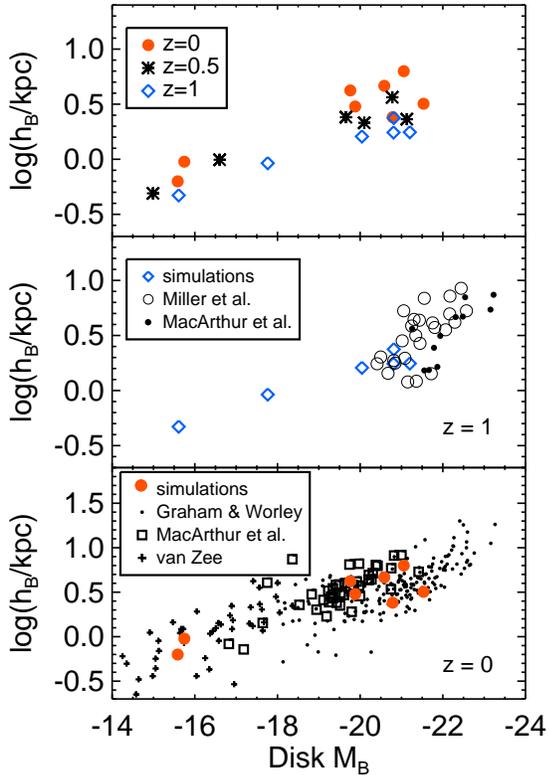} \caption{Rest frame $B$ band disk scale length as 
a function of magnitude for our simulated galaxies.  {\it Top panel}: The 
evolution of the population with time.  {\it Middle panel}: Simulated points 
at $z$=1 as in the top panel, but now compared to observational results 
for decomposed disks from \citet{macarthur08} and Miller et al.\,({\it in 
prep}).  {\it Bottom panel}: Same as Fig.~\ref{fig:z0}, shown again for 
comparison to the high $z$ results. }
\label{fig:magsize}
\end{figure}

The evolution of the magnitude -- size relation for these simulated 
galaxies is shown in Fig.~\ref{fig:magsize}.  The top panel shows the 
simulation results at $z$=0, 0.5, and 1.  It is evident that the magnitude -- 
size relation for these galaxies is evolving in time, dimming in surface 
brightness since $z$=1.  Despite the fact that these are individual 
galaxies being followed in time, the snapshot of the population of disks 
at each redshift is consistent with the observations.  This is evident in 
the bottom two panels.  The middle panel compares the simulated disk scale 
lengths at $z$=1 to the decomposed $i$ band disk scale lengths of galaxies 
at 0.9 $< z < $ 1.2 from \citet{macarthur08} and Miller et al.\,({\it in 
prep}).  No dust corrections are applied to the observational $z$=1 data, 
as these galaxies are expected to be low metallicity where corrections 
are uncertain and often ignored \citep{dutton10b}.  Any realistic dust 
corrections are likely to be tiny, and not affect the results.
The bottom panel is a reproduction of Fig.~\ref{fig:z0}, shown again for 
easy comparison of the results across redshifts.  While observational 
biases allow for a direct comparison only at the massive end at $z$=1, the 
disk sizes of the simulated galaxies are in good agreement with observed 
disk sizes, as a function of redshift.  We note that observational data 
that used a single Sersic component fit \citep[e.g.,][]{barden05} are also 
in good agreement with the $z$=1 data presented here (and lie in the same 
magnitude range), despite the lack of a bulge/disk decomposition.  For 
clarity, we show only the decomposed data in Fig.~\ref{fig:magsize}. 

\begin{figure}
\plotone{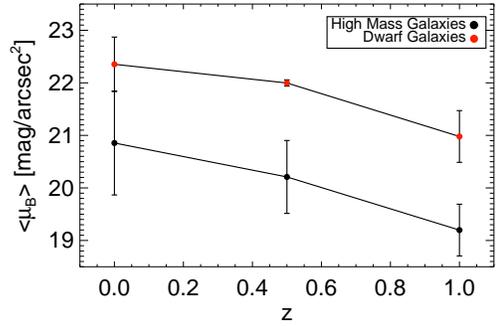}
\caption{The evolution of surface brightness as a function of redshift.  
The two lowest mass galaxies are shown separately from the high mass 
galaxies.  Data points at each $z$ show the mean of the population, and 
the error bars reflect the standard deviation of the sample.  } 
\label{deltamu}
\end{figure}

The top panel of Fig.~\ref{fig:magsize} shows that there has been apparent 
dimming of the surface brightness of the simulated population since $z$=1. 
Fig.~\ref{deltamu} quantifies this evolution, showing the change in the 
surface brightness, $\mu_B$, of the simulated galaxy disks.  We have defined 
$\mu_B$ such that
\begin{equation}
\mu_B = M_{B,disk} + 5 log h_B + 2.5 log (2\pi).
\label{eq:1}
\end{equation}
Data points 
in Fig.~\ref{deltamu} at $z$=0, 0.5 and 1 are for the mean $\mu_B$ for the 
population at that time, and error bars reflect the standard deviation of 
the sample.   Because our two dwarf galaxies are too faint to be generally 
observable at $z$=1, Fig.~\ref{deltamu} shows the evolution of these two 
galaxies separately from the higher mass galaxies.  However, the evolution 
is roughly similar in both mass ranges, with about 1.5 magnitudes in 
surface brightness dimming between $z$=1 and $z$=0.  A dimming of 1.5 
magnitudes is consistent with observational studies that have examined the 
surface brightness evolution of disk dominated galaxies 
\citep[e.g.,][]{simard99, rav04, trujillo04, barden05, melbourne07, 
kanwar08}.  

\subsection{Evolution of Individual Galaxies}

\begin{deluxetable*}{lcccccccc}
\tablecaption{Evolution of Galaxy Disk Properties \label{evol} }
\tablewidth{0pt}
\tablehead{\colhead{simulation} &
\colhead{$\mu_{B,0}$} & \colhead{$\Delta \mu_{B,0}$} &
\colhead{M$_{B}$} & \colhead{$\Delta$M$_{B}$} & 
\colhead{h$_{B}$} & \colhead{$\Delta$h$_{B}$} &
\colhead{M$_{*}$} & \colhead{$\Delta$M$_{*}$} \\
  & \colhead{mag/$\square$''} & \colhead{$\Delta$mag/$\square$''} & \colhead{mag} & \colhead{$\Delta$mag} & 
\colhead{kpc} & \colhead{$z$=0/$z$=1} & \colhead{\Msun} & \colhead{$z$=0/$z$=1} \\
  & \colhead{$z$=0} & \colhead{to $z$=1} & \colhead{$z$=0} & \colhead{to $z$=1} & 
\colhead{$z$=0} & \colhead{   } & \colhead{$z$=0} & \colhead{  } \\
\colhead{(1)} & \colhead{(2)} & \colhead{(3)} & \colhead{(4)} & 
\colhead{(5)} & \colhead{(6)} & \colhead{(7)} & \colhead{(8)} & 
\colhead{(9)} } 
\tabletypesize{\footnotesize}
\startdata
h516 & 22.7 & 1.5 & -15.7 & 2.0 & 0.95 & 1.0 & 4.3$\times$10$^8$ & 1.69 \\
h799 & 22.1 & 0.7 & -15.6 & 0.0 & 0.6 & 1.3 & 1.8$\times$10$^8$ & 1.32 \\  
h603 & 22.5 & 2.8 & -19.8 & 1.0 & 4.2 & 2.4 & 1.3$\times$10$^{10}$ & 2.78 \\  
h986 & 21.2 & 1.0 & -19.9 & 0.2 & 3.0 & 1.9 & 1.0$\times$10$^{10}$ & 3.22 \\  
h239\tablenotemark{a} & 19.3 & N/A & -21.5 & N/A  & 3.2 & N/A & 9.0$\times$10$^9$ & 10.17 \\
h258\tablenotemark{a} & 21.3 & 1.6 & -21.1 & -0.2 & 6.3 & 2.7 & 3.2$\times$10$^{10}$ & 2.53 \\ 
h277 & 19.7 & 0.6 & -20.8 & 0.4 & 2.4 & 1.4 & 1.9$\times$10$^{10}$ & 1.95 \\  
h285 & 21.4 & N/A & -20.6 & N/A & 4.6 & N/A & 2.0$\times$10$^{10}$ & 3.07 \\  
\enddata
\tablenotetext{a}{Measured at z=1.25; undergoing major merger at z=1.}
\tablecomments{For columns (3) and (5) ($\Delta \mu_{B,0}$ and $\Delta$M$_{B}$), 
positive values represent dimming between $z$=1 and $z$=0.  A negative 
value means that the galaxy disk is brighter at $z$=0.  Note that values 
listed in this table are for the disk only, while those in Table~\ref{obssum} 
are for the entire galaxy (disk and spheroid).  h239 and h285 do not have 
exponential disks at $z$=1.  The stellar disk masses have been measured 
directly from the star particles in the simulations, based on a kinematic 
disk decomposition.  See Section~\ref{starmass} for details.  }
\end{deluxetable*}

The previous section examined the population of simulated disks in terms 
of surface brightness evolution.  We now decompose the surface brightness 
trends into luminosity and size evolution, and demonstrate how individual 
galaxies evolve in each property with time.  Table~\ref{evol} quantifies 
the changes in the disk $B$ band central surface brightness ($\mu_{B,0}$, 
as opposed to $\mu_{B}$ plotted in Fig.~\ref{deltamu}), size, and magnitude 
for each of our simulated galaxies.

\begin{figure*}
\plotone{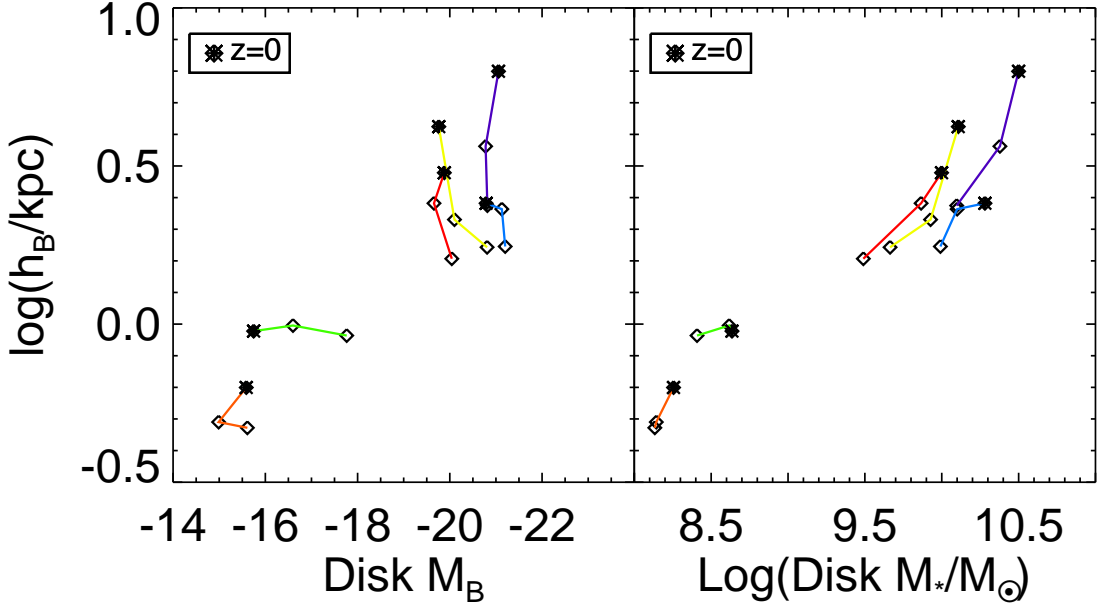}
\caption{{\it Left panel}: Evolution in the magnitude -- size plane with time.  
Each color connects an individual galaxy at $z$=1, 0.5, and 0.  The higher 
$z$ steps are marked by diamonds, and the $z$=0 step is indicated by a cross 
within the diamond.  
Note that low mass galaxies evolve more in luminosity (dimming), while higher 
mass galaxies evolve more in size.  {\it Right panel}: Evolution in the 
stellar mass -- size plane as a function of time.  Note that these galaxies 
grow so that they remain on approximately the same relation with redshift. }
\label{dmagsize}
\end{figure*}

\subsubsection{Luminosity Dimming since $z$ = 1}

It is widely expected that galaxies should undergo dimming since 
$z$=1, due to declining star formation rates. The key question is 
how much galaxies dim.  It is already apparent from Fig.~\ref{fig:magsize} 
that our more massive galaxies are undergoing growth in size between 
$z$=1 and $z$=0, meaning that the change in surface brightness of 
$\sim$1.5 magnitudes over this time cannot be entirely due to 
luminosity dimming at a fixed size.  
The changes in size and magnitude that are quantified in Table~\ref{evol} 
are also visualized in Fig.~\ref{dmagsize}.  Now it can be seen that 
there is in fact a dependency on mass in the evolution of 
the galaxies.  Our more massive galaxies are growing in size and generally 
undergoing only a small amount of dimming in total magnitude; some 
are even getting brighter with time.  Meanwhile, there is little 
change in the scale length of our lowest mass galaxies, but their 
evolution in luminosity is erratic (for this sample of two galaxies) 
due to their bursting star formation histories.  

The luminosity evolution of the low mass dwarf galaxies is tied up 
with their particular star formation histories.  While h799 is quiescent 
between $z$=1 and $z$=0.5 so that it dims, it undergoes another burst 
of SF at low $z$, increasing its magnitude.  On the other hand, h516 
stays relatively quiescent between $z$=1 and $z$=0, so that it 
undergoes significant luminosity dimming (of 2.0 magnitudes).  This 
large amount of dimming may at first seem to be at odds with the fact 
that low mass galaxies are generally undergoing more SF today than they 
were in the past (this is discussed further in Section~\ref{starmass}, 
and is one form of galaxy ``downsizing'').  However, this extreme 
dimming for low mass galaxies has been seen in observational 
results as well \citep{melbourne07, kanwar08}.  \citet{melbourne07} 
concluded that small galaxies (those with half light radii\footnote{ 
The half light radius is a factor 1.68 times larger than the disk scale 
length for an exponential disk.} $\lesssim$ 2 kpc) have undergone 
roughly 2.5 magnitudes of dimming since $z$=1.  Given the bursty 
SF nature of the two low mass dwarfs we have in our sample, it is not 
possible to compare statistically to the observations, but there is 
nothing to suggest that our simulated dwarfs are inconsistent with the 
\citet{melbourne07} results. 

At the higher mass end, \citet{melbourne07} find $\sim$1.5 mag of 
dimming since $z$=1 for galaxies with half light radii $>$ 3 kpc 
(disk scale lengths $>$ 1.8 kpc), if the strictest interpretation is 
made that all surface brightness evolution is in luminosity 
rather than size.  Clearly, this is not true for our high mass 
galaxies, which are growing with time.  

If galaxies are growing with time, then comparing galaxies that are 
a given size (e.g., 3 kpc) at $z$=0 with galaxies that are of similar 
size at $z$=1 means that similar galaxies are {\it not} being compared.  
Those $z$=0 galaxies will have been smaller at $z$=1, and it is unknown 
from our simulation results what size galaxies with scale lengths 
larger than $\sim$2.5 kpc at $z$=1 will 
have evolved to by $z$=0.  In summary, if galaxies are growing since 
$z$=1, this will mimic a larger change in luminosity on the magnitude -- 
size relation than actually occurs \citep{trujillo04, barden05, 
trujillo05}.

\subsubsection{Change in Size since $z$ = 1}

\begin{figure}
\plotone{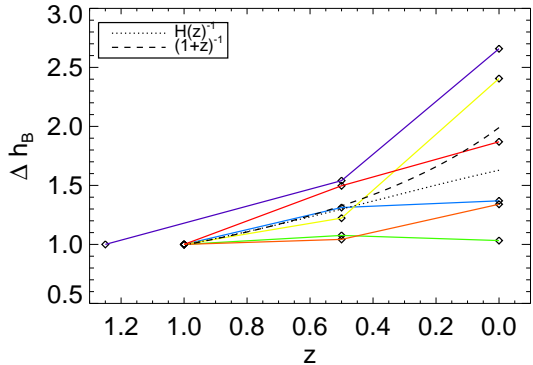}
\caption{The growth of the simulated disks, normalized to their high $z$ 
sizes.  The dashed and dotted lines show simple predictions for the growth 
of disk galaxy populations since $z$=1.  The colors here are the same 
for individual galaxies as in Fig.~\ref{dmagsize}. }
\label{evolsize}
\end{figure}

Our simulated disk galaxies {\it as a population} appear to match the 
observational results for the evolution in the size-magnitude plane very well 
(Fig.~\ref{fig:magsize}).  Yet Fig.~\ref{dmagsize} demonstrates what the 
observations cannot witness: individual galaxies are growing with time, 
so that the evolution in the magnitude -- size plane cannot be due to 
dimming alone.  Disk scale length evolution for each galaxy is plotted 
in Fig.~\ref{evolsize}, normalized to their size at $z$=1.  These values 
are also quantified in Table~\ref{evol}.  Only a few of these galaxies show 
negligible size evolution, with the two lowest mass galaxies undergoing 
the smallest size changes.  The remaining galaxies show clear growth, and 
about half evolve by the amount predicted by the simple SIS model, (nearly a 
factor of two back to $z$=1), or more.  The dotted line in Fig.~\ref{evolsize} 
shows the growth predicted for the simple SIS model \citep{mmw}, while the 
dashed line shows the growth determined by \citet{mao98} using 16 galaxies 
at $z$=1.  

However, it is not necessarily expected that any given galaxy should 
follow the $H(z)^{-1}$ analytic growth predicted by the SIS model. 
\citet{somerville08} demonstrated that more sophisticated models adopting 
NFW profiles could bring the observations and theory closer in line.  
The prediction that sizes scale with $H(z)^{-1}$ is based on the assumption 
that
\begin{equation}
r_{200} = V_{c}/10 H(z)
\label{eq:2}
\end{equation}
where $V_c$ is the circular velocity of the halo and $r_{200}$ is the radius 
at which the mean density is equal to 200 times the critical density.  Then 
\begin{equation}
h \propto \lambda r_{200}
\label{eq:3}
\end{equation}
where $\lambda$ is the dimensionless spin parameter \citep{peebles69}.
If the disk size follows $H(z)^{-1}$, both $\lambda$ and $V_c$ must stay roughly 
constant with time.  While $V_c$ does stay roughly constant back to $z$=1 
in these galaxies, $\lambda$ does not.  It is known that $\lambda$ may vary 
for individual halos due to mergers or smooth accretion \citep{gardner01, 
maller02a, vitvitska, peirani04, hetznecker06, dn07}.  The previous work of 
analytic models do not require $\lambda$ to stay constant, as they simply 
use a snapshot in time and the spin values of halos at that snapshot.  That 
is, they study a population of galaxies at a given time, and make no attempt 
to follow an individual galaxy with time as we do here.
Hence, it is not surprising that some of the simulated galaxies presented 
here deviate from growing precisely as $H(z)^{-1}$.

A better question, if we want to compare our disk sizes to analytic models, 
is to ask if the scale lengths measured based on our light profiles are 
similar to those predicted by analytic models.  For an NFW density 
distribution, the disk scale length, R$_d$ is  
\begin{equation}
R_d = \frac{1}{\sqrt{2f_{c}}}\frac{j_{d}\lambda}{m_{d}}r_{200} f_{R}(\lambda,c,m_d,j_d)   
\label{eq:4}
\end{equation}
where $c \equiv r_{200}/r_{s}$ measures the halo concentration,
and $r_s$ is NFW the scale radius.  $f_c$ uses $c$ to determine the 
energy of the NFW halo compared to the SIS model, and $f_{R}(\lambda,c,m_d,j_d)$ 
is a factor that accounts for the gravitational effect of the disk (adiabatic 
contraction).  $j_d = J_{disk}/J_{200}$ and $m_d = M_{disk}/M_{200}$ are the 
fraction of disk angular momentum and mass, respectively, to the total halo.  
For sizes to scale as $H(z)^{-1}$, it is assumed that $j_d/m_d$ = 1.  This is 
likely to not be true, though, as demonstrated by \citet{sales09}.  

We have determined the values of $f_c$, $f_R$, $m_d$, and $j_d$ for each 
of our galaxy halos at $z$=0, directly from the simulation results.   We 
compared the $R_d$ expected from these quantities, as derived by equation 
\ref{eq:4}, to the $i$ band scale lengths listed in 
Table~\ref{obssum}.  The results agree to within 10\% for only the two 
lowest mass galaxies.  For the more massive galaxies, the predicted NFW 
$R_d$ is consistently shorter than that found for the $i$ band, and for 
four of the six massive galaxies, the measured $i$ band results are a 
factor of 2-4 larger than the NFW prediction. 

The main source of this discrepancy appears to be in the value obtained 
for the spin value, $\lambda$, for the entire halo.  If the spin of the gas, 
$\lambda_g$, within the halo is used instead, the discrepancy shrinks so 
that six of the eight galaxies have NFW $R_d$ and $i$ band results within 
10\% of each other.\footnote{We note that the two that remain discrepant 
both show evidence for a downward ``break'' in the exponential light profile   
at large radii.}  As listed in Table~\ref{simsum}, the values of $\lambda$ 
are generally lower than $\lambda_g$ for all halos except the two lowest 
mass galaxies.  This trend for $\lambda_g$ to be larger than $\lambda$ 
has been seen previously in simulations \citep{sharma05}, and has been 
suggested as one reason why disks may have a smaller fraction of low 
angular momentum than their dark matter halos \citep{chen03}.  Recently, 
\citet{roskar10} used a similar simulation run with {\sc Gasoline} to 
show that gas that enters the virial radius of a halo and cools toward 
the disk is torqued by the hot halo gas, so that the angular momentum of 
the disk gas becomes aligned with the hot halo.  This is true even for 
gas that is initially counter-rotating with the disk, and hence is a 
powerful method to remove negative and low angular momentum material 
from the disk.  
The lack of low angular momentum material for the disk gas will lead to 
larger values of $\lambda_g$ compared to $\lambda$ for the total halo.  
The disk stars, whose light we trace in the results of this paper, form 
from this cold gas with larger spin values, and thus $\lambda_g$ will be 
a better predictor for the NFW $R_d$ than $\lambda$.  
 
The result that the disk baryons are lacking a low angular momentum 
component compared to the DM is not trivial.  This result needs to be 
examined in detail.  However, this requires a full evaluation of the 
history of angular momentum in these halos, which is beyond the scope of 
the current paper.  We reserve such a study for future work.

\section{Growth of the Stellar Disk since $z$ = 1}
\label{starmass}

As discussed above, numerous observational studies have found that there 
has been a decline in the surface brightness of disk galaxies, by a 
magnitude or more, since $z$=1.  Due to the fact that there is little 
evidence for a change in the size function of disk galaxies over this 
time period \citep[e.g.,][]{kanwar08}, the observations can only be 
interpreted as an upper limit to luminosity evolution.  That is, if 
disks do not change in size, then all of the evolution must be in 
luminosity.  However, due to the degeneracy in the magnitude -- size 
plane, the option of possible growth is left open.  

\citet{barden05} investigated this disk growth.  After showing that they, 
too, agreed with $\sim$1 magnitude of surface brightness dimming back to 
$z$=1, they then used galaxy colors to derive stellar masses and considered 
the stellar mass -- size relation for their galaxies.  They found that this 
relation, unlike the magnitude -- size relation, showed little or no 
evolution back to to $z$=1. 
Disk growth may still occur, but must occur in such a way that galaxies remain 
on the same stellar mass -- size relation with time.  They concluded that this 
was evidence for weak inside-out growth of galaxy disks, as did additional 
later studies \citep{trujillo05, trujillo06, dutton10b}.  

The right panel of Fig.~\ref{dmagsize} shows the evolution in the stellar 
mass -- size plane for our simulated galaxies.  In agreement with previous 
results \citep{barden05, brook06, far09}, these galaxies are growing along a path 
that keeps them on the same stellar mass -- size relation with redshift.  

The stellar mass shown in the right panel of Fig.~\ref{dmagsize} is for 
the disk only.  To separate the disk growth from total stellar growth of 
these galaxies (i.e., from the bulge and halo spheroidal stellar components), 
a kinematic decomposition was done to identify disk stars at $z$=0.  To 
identify disk particles, the galaxies are first aligned so that the disk 
angular momentum vector lies along the {\it z}-axis.  J$_z$/J$_{circ}$
is calculated for each star particle in the galaxy, where J$_z$ is
the angular momentum in the {\it x-y} plane, and J$_{circ}$ is the momentum
that a particle would have in a circular orbit with the same orbital energy.
Disk stars are identified as those having near circular orbits, so that 
J$_z$/J$_{circ}$ $>$ 0.8.\footnote{This criterion corresponds to an 
eccentricity $\leq$ 0.2, which matches the eccentricities observed 
in the Milky Way disk \citep{nordstrom04}.}
These $z$=0 disk stars were then searched for in their most massive 
progenitor at $z$=0.5 and $z$=1 to find the mass of these stars that had 
formed at each step, with the results plotted in Fig.~\ref{dmagsize}.  

\begin{figure}
\plotone{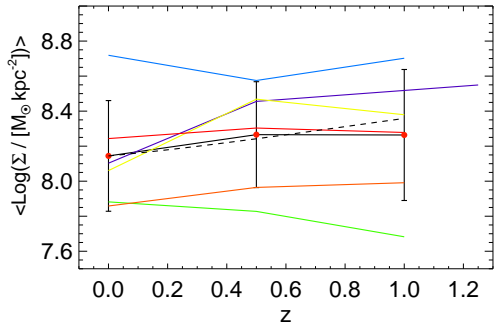}
\caption{The surface mass density of the simulated disks as a function of 
redshift.  The red points with error bars, connected by the solid black 
line, show the mean and standard deviation of the sample of nine at 
$z$ = 0, 0.5, and 1.  The mean varies by only 0.1 dex between $z$=1 and 
$z$=0.  The dashed line shows the analytic prediction for the SIS model. 
The colored lines show the evolution of each individual disk galaxy.  Colors 
are the same for individual galaxies as in Figs.~\ref{dmagsize} and 
\ref{evolsize}.
 }
\label{barden}
\end{figure}

If galaxy disks are evolving approximately along the same stellar mass -- 
size relation with time, then the surface mass density of disks should 
show little evolution back to $z$=1 \citep{barden05, somerville08}.  That 
is, a galaxy with stellar mass of 10$^{10}$ \Msun ~at $z$=1 is roughly 
the same size (though maybe just slightly smaller) as a galaxy at $z$=0 
with 10$^{10}$ \Msun, yielding roughly the same surface densities.  
Fig.~\ref{barden} tentatively confirms this result, showing that the 
surface mass density of our population of disks evolves little with time, 
with surface mass density, $\Sigma$ defined as
\begin{equation}
log \Sigma = log M_{*,disk} - 2 log h_B - log (2\pi) .
\label{eq:5}
\end{equation}
The red points with error bars, connected by the solid black line, show 
the mean and standard deviation for the whole sample at $z$ = 
0, 0.5, and 1.  The colored lines are the results for individual 
simulated disks with time.  We note, however, that the dashed line shows 
the prediction for the surface mass density evolution for the simple 
SIS model \citep{mmw}.  
The standard deviation within our small sample is large, so that the 
simulated galaxies are fully consistent with the predicted growth.  
While our sample is clearly too small to derive statistical results,  
the change in the mean surface density of the population between 
$z$=1 and $z$=0 is only 0.1 dex, half of the evolution predicted by 
the SIS model.  We note that an updated model by \citet{somerville08} 
that incorporates the evolution of NFW halos derived in N-Body 
simulations predicts only $\sim$0.2 dex change in the stellar surface 
mass density back to $z$=1.  This small evolution in surface mass 
density is due to evolution along a stellar mass -- size relation 
that changes little with time. 

\subsection{The SFR -- Mass Relation}

\begin{figure*}
\plotone{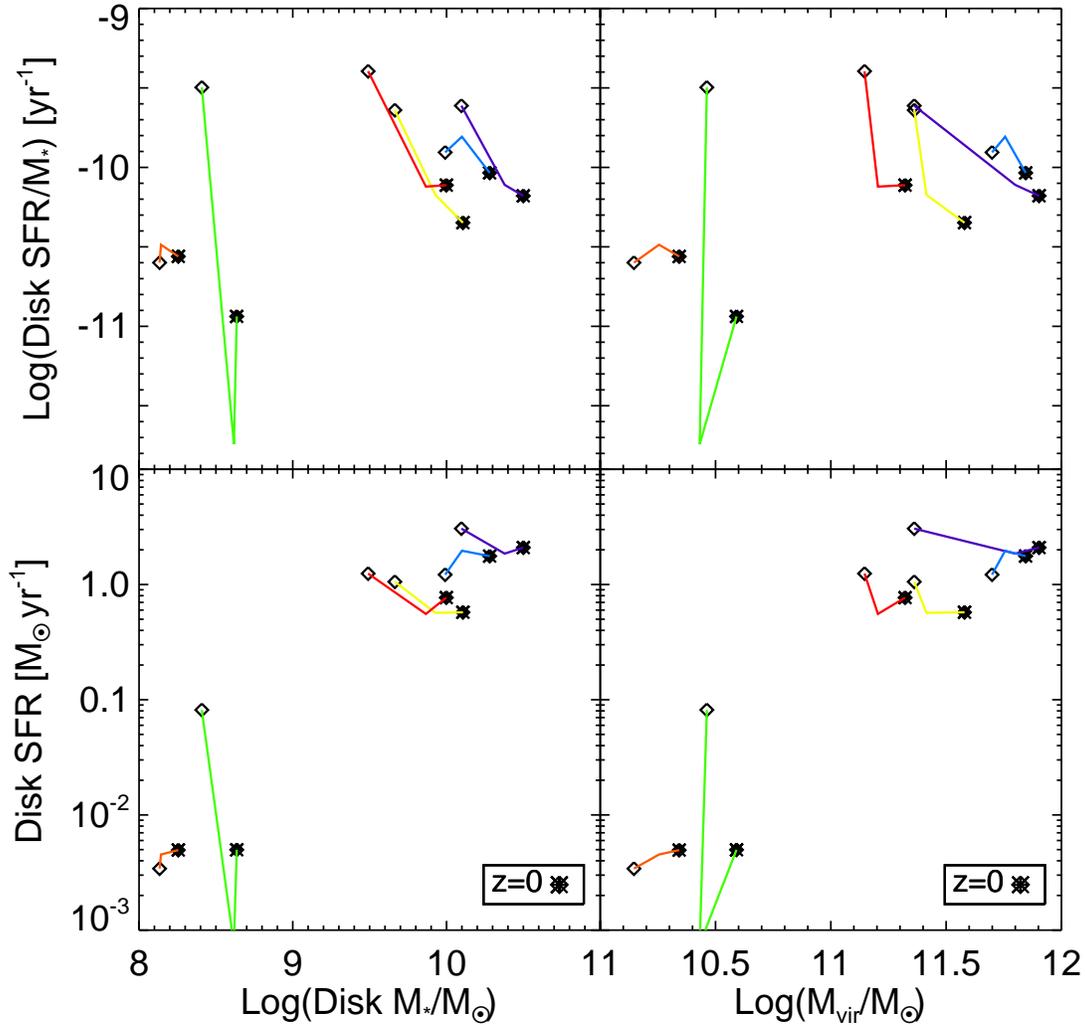}
\caption{The SFR and sSFR of simulated galaxies.  {\it Left panels}: The 
bottom panel shows the SFR in the disk of each galaxy as a function of 
the stellar disk mass.  The top panel is the specific SFR of the disk.
{\it Right panels}: Same as the left panels, but now in terms of the  
total virial mass of the halo.  The colors here are the same for individual 
galaxies as in previous figures.  }
\label{fig9}
\end{figure*}

Due to the fact that the stellar mass -- size relation shows little change 
with time, examining the evolution of galaxy disks at a fixed {\it stellar 
mass} is similar to examining the evolution at a fixed {\it size}. 
However, because galaxies are growing with time, a galaxy that is, e.g., 
3 kpc in size and  10$^{10}$ \Msun ~in stellar mass at $z$=1 has increased 
in mass and size by $z$=0 and another galaxy has moved from a lower mass and 
smaller size into the same bin.  Observations, and the results presented 
here, tell us that these two galaxies of same size at $z$=1 and $z$=0 
have a different luminosity: a galaxy at $z$=1 with similar stellar mass 
and size to a $z$=0 galaxy will be brighter.  Note that this is {\it not} 
equivalent to saying that an individual galaxy was brighter in the past.
Instead, galaxies at fixed stellar mass were brighter in the past.  Because 
they are the same stellar mass, the brighter $z$=1 galaxy must then be forming 
more stars.  

Two of our galaxies can be compared in this way.  Galaxy h603 has a stellar 
disk mass of 1.28$\times$10$^{10}$ \Msun ~at $z$=0, while h258 has a stellar 
disk mass of 1.25$\times$10$^{10}$ \Msun ~at $z$=1.  The SFR within the last 
100Myr for h258 at $z$=1 is 3.1 \Msun/yr, while it is 0.6 \Msun/yr for h603 
at $z$=0.  The lower SFR for h603 is due to the fact that it resides in a 
less massive halo than h258 (see Table~\ref{simsum}).  Even at $z$=1, h603 
had a lower SFR than h258, of 2.1 \Msun/yr.  This is because these galaxies 
follow a SFR -- stellar mass relation \citep[which also leads to their 
reproduction of the observed stellar mass -- metallicity relation,][]{Brooks07}.  

Fig.~\ref{fig9} shows the SFR -- stellar mass and specific SFR (sSFR) -- stellar 
mass relations for our simulated disks.  The left panels of Fig.~\ref{fig9} 
shows the sSFR (top) and SFR within the last 100 Myr (bottom) versus 
stellar disk mass.  These left panels confirm the values listed in the 
paragraph above.  The dwarf galaxies are included in this plot for full 
disclosure, but lie outside of the mass range observable at high $z$, unlike 
the higher mass galaxies.  Two trends among the higher mass galaxies are 
noteworthy.  First, the SFR of any individual galaxy tends to decrease 
slightly with time, but second, the entire population as a whole can be 
observed to shift to lower SFRs and sSFRs \citep{guo08, dutton10a, bouche10}.  
The decline in SFR and sSFR in has been observed extensively in 
populations of galaxies \citep{gavazzi96, boselli01, 
K03, brinchmann04, feulner05, Erb06, salim07, noeske07a, noeske07b, 
elbaz07, Daddi07, schiminovich07, cb08, pannella09, damen09a, damen09b, 
dunne09, rodighiero10, oliver10, mannucci10, laralopez10}, and is one 
form of ``downsizing'' in galaxy evolution.

The right panels of Fig.~\ref{fig9} also show SFR and sSFR for the 
simulated galaxy disks, but now as a function of total halo mass.  This 
is included to demonstrate that although, e.g., h603 at $z$=0 and h258 
at $z$=1 have similar disk stellar masses, this does not correspond to 
similar virial masses for those redshifts (colors yellow and purple in the 
figures, respectively).  The deeper potential well of the more massive 
galaxy (h258) leads to a higher SFR.\footnote{Note that although these 
two galaxies appear to have the same virial mass at the initial step, 
the properties of h258 have actually been measured at $z$=1.25 due to 
the fact that it is undergoing a merger at $z$=1.  Galaxy h258 is 50\% 
more massive than h603 at $z$=1. } 

In conclusion, attempting to compare galaxies that are a similar stellar 
mass at $z$=1 and $z$=0 (and by the tight stellar mass -- size relation, 
similar size) leads to inherently comparing galaxies of two different 
virial masses.  This then leads to the observed difference in luminosity 
for galaxies of a similar size at $z$=1 and $z$=0, caused by the SFR -- 
stellar mass relation, and the corresponding stellar mass -- halo mass 
relation \citep{guo09, moster10}.

\subsection{Comparison to Observed Relations}
\label{comp}

\begin{figure*}
\plotone{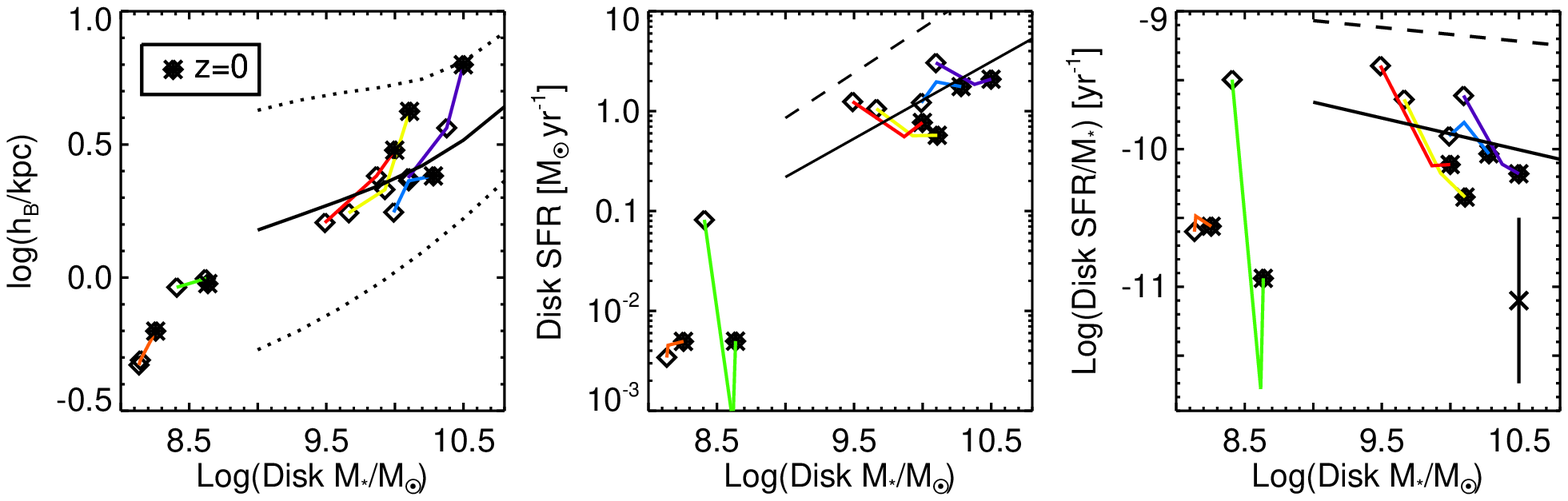}
\caption{Comparison of simulated disk galaxy scaling relations to 
observational results. {\it Left}: The stellar mass -- 
size relation, reproduced from the right panel of Fig.~\ref{dmagsize}, now 
showing (solid line) the $z$=0 observed relation of \citet{dutton10b} plus 
the 2$\sigma$ scatter in the observations (dotted lines).  Their observed 
 $z$=1 relation would simply be shifted downward 0.1 dex in size.  
{\it Center}: The 
stellar mass -- SFR relation, as in Fig.~\ref{fig9}, now showing the 
observed relations from \citet{elbaz07} at $z$=0 (solid line) and at $z$=1 
(dashed line).  {\it Right}: As for the center plot, but now for the sSFR.  
The x data point with error bars in the lower right corner of the panel 
represents roughly the 2$\sigma$ scatter in the \citet{elbaz07} data. }
\label{fig10}
\end{figure*}

In Fig.~\ref{fig10} we reproduce some of the simulated scaling relations, 
this time with observed scaling relations directly overplotted for 
comparison.  The left panel of Fig.~\ref{fig10} shows the low redshift 
relations for SDSS galaxies (solid line, with 2$\sigma$ scatter in dotted 
lines) from \citet{dutton10b}.  It is worth noting that \citet{dutton10b} 
performed bulge/disk decompositions with the SDSS data, so that the sizes 
plotted are fits to purely exponential disk scale lengths (converted 
from the half light radius presented in their paper).  \citet{dutton10b} 
also use higher $z$ data from the DEEP2 survey to measure the change in 
size since $z$=0, finding $\sim$0.1 dex of size evolution at a fixed 
mass back to $z$=1.  In the mass range of overlap (M$_*$ $>$ 10$^9$ \Msun), 
our simulated galaxies are entirely consistent with observed galaxies 
within the scatter. 

The central panel of Fig.~\ref{fig10} reproduces the stellar mass -- 
SFR relation, now with observational data from \citet{elbaz07} plotted  
for $z$=0 (solid line) and $z$=1 (dashed line), corrected from a 
Salpeter IMF to a Kroupa IMF.  At both redshifts, the minimum mass 
observed is roughly 10$^9$ \Msun.  The right panel shows the same 
data, but now for the sSFR.  The data point with error bars 
in the lower right corner of the right panel provides an estimate of 
the typical scatter in the \citet{elbaz07} data.  While our galaxies 
are certainly consistent with the data within the scatter, and the 
general slope at each $z$ is consistent with the observations, there 
is overall a tendency for the simulated galaxies to be more massive 
than the observations.  We believe the discrepancy is in stellar mass 
rather than SFR, because these simulated galaxies have both $z$=1 and 
$z$=0 disk SFRs in good agreement with observed galaxies in the same 
halo mass range \citep{cw09, boissier10}.
We note that the \citet{elbaz07} data is for total stellar mass, while 
our simulated results are for the disk only (bulge excluded).   
However, including the bulge mass will only exacerbate the discrepancy. 
For the massive galaxies, the discrepancy is probably due to the fact 
that these simulations are known to form too many stars \citep{guo09, 
moster10}.  

We can learn something about {\it when} this overproduction of stars 
occurs in the simulations by also examining the stellar mass -- sSFR 
plot (right panel of Fig.~\ref{fig10}).  While within the observed 
scatter, is appears that the sSFRs of the simulated disks are 
consistently lower than observational results at both $z$=1 and $z$=0 
\citep{elbaz07, noeske07a, salim07}.  These plots demonstrate that 
these simulated galaxies must be forming too many stars at early times, 
at redshifts higher than plotted here ($z >$ 1).  Because the SFRs of 
the galaxies are in good agreement with observed galaxies in the same 
halo mass range at both $z$=1 and $z$=0 \citep{cw09}, the fact 
that the sSFRs are lower than observed values requires that the stellar 
mass of these disks be too large by $z$=1.

The story is slightly different for the two dwarf disk simulations.  
These two dwarf galaxies have been published previously in 
\citet{bulgeless}, where it was demonstrated that their properties 
are a good match to observed dwarf disk galaxies \citep[see also]
[]{sanchezjanssen10}, including stellar mass \citep{vdb01}.  However, 
the observed sSFRs of low mass galaxies is on average higher than that 
of massive galaxies \citep{salim07},
indicating that low mass galaxies have been very inefficient in the 
past at turning gas into stars.  Clearly, that is not the case with our 
two simulated dwarf disk galaxies, which have too low sSFRs compared 
to most observed dwarf galaxies due to the fact that they form the bulk 
of their stars prior to $z$=1.  Hence, while approximately the correct 
mass in stars is formed, the star formation is biased toward $z >$ 1.
We have verified that these two simulations, when run with stronger SN 
feedback, have burstier SF histories with their SF spread out more 
evenly down to the present time.   This leads to approximately the same 
mass in stars formed by $z$=0, but with a more uniform SF history.  
This means that the stronger feedback dwarf runs have higher sSFRs than 
currently presented here, and sSFR values larger than the massive 
galaxy simulations shown here.  
Future work will include these new dwarf disks, 
which show similar stellar masses, rotational velocities, and sizes 
to those presented in this paper.  

Thus, in all of the simulated galaxies the star formation history is 
biased toward too much star formation at high $z$.  By $z$=0, though, 
the dwarf galaxies have formed approximately the observed amount of stars 
for their halo mass, while the more massive galaxies have overproduced 
the total amount of stars.  It is suspected that this overproduction of 
stars in the massive galaxy disks may be the result of the adopted low 
SF density threshold, which allows for most of the disk to be capable of 
SF at any given time, unlike the high density threshold which limits SF 
only to rare density peaks.  However, as discussed in Section 2.1, it is 
currently computationally expensive to generate the more massive galaxies 
with the force resolution required to resolve high density peaks where 
SF occurs, and no simulations of sufficiently high resolution are 
currently available to test this hypothesis.  
The adopted SF threshold is a compromise that allows stars to form at 
radii observed in real galaxy disks, and we have demonstrated that it 
leads to scale lengths in good agreement with observed galaxies.  
However, we note that even if our massive galaxies are forming too many 
stars, it is by roughly a factor of 3 at most, and at early times.  
This means that the more massive galaxies may shift up to 0.5 dex to 
lower stellar masses in the stellar mass -- size plane, but will still 
grow along the stellar mass -- size relation, so that our conclusions 
are unaffected.  Likewise, the $B$ band magnitudes we present here are 
unlikely to be affected, as the bulk of overproduced stars must occur 
at $z >$ 1.  Since $B$ band light is dominated by young stars, and our 
low $z$ SFRs are in good agreement with observed galaxies in the same 
halo mass range, the magnitude -- size relations presented in Figs.
\ref{fig:z0} and \ref{fig:magsize} are unlikely to change.  Hence, the 
conclusions presented here about the origin of the evolution in the 
size -- luminosity relation remain valid.

\section{Conclusions}

For the first time, a sample of very high resolution galaxy disks 
simulated within a fully cosmological context have been used to determine 
the evolution of the galaxy disk size -- magnitude relation.  Unlike 
previous simulations, we determine disk scale lengths and magnitudes by 
fitting the stellar light profile rather than decomposing our simulations 
based on kinematics (and hence mass).
We generate artificial surface brightness images using {\sc Sunrise} 
\citep{sunrise, sunrise2}, in order to derive disk properties using 
methods used by observers.  Exponential disk scale lengths were derived 
using {\sc Galfit} \citep{galfit} with redshifted, surface brightness 
dimmed $i$ band images at $z$=1 and 0.5, and rest frame $B$ band images 
at $z$=0.  This allows for a direct comparison between simulated and 
observational results in the size -- magnitude plane.  

Fig.~\ref{fig:z0} demonstrates that these galaxies 
overcome past inabilities for cosmological disk simulations to retain 
angular momentum.  Artificial loss of angular momentum is minimized due 
to high resolution and a physically motivated SN feedback recipe, and 
hence we produce disks with realistic sizes compared to observed disks.  
Fig.~\ref{fig:magsize} shows that the sizes are in agreement with 
observed disk sizes back to $z$=1.  This success, and our previous 
successes in matching the observed stellar mass -- metallicity relation 
with time \citep{Brooks07}, and the Tully-Fisher relationship 
\citep{merger}, indicate that these simulated disks reproduce the 
fundamental scaling relations for disk galaxies.  Having established 
this, we can now use these simulated disks to investigate and interpret 
observational findings.

We use the simulated disk properties to investigate the evolution in 
the magnitude -- size relation with time.  Observations have generally 
concluded that there has been 1 to 1.5 magnitudes of surface brightness 
dimming since $z$=1 \citep{schade96, roche98, lilly98, bs02, rav04, 
trujillo05, barden05, melbourne07, kanwar08}.  We find similar evolution 
for our galaxies (see Fig.~\ref{deltamu}).  Combined with a lack of 
evidence for an evolving size function for galaxy disks \citep{lilly98, 
rav04, kanwar08}, and little to no growth in galaxy disks sizes back 
to $z$=1 \citep{barden05, trujillo05}, the observed surface brightness 
evolution can only be interpreted as an upper limit to luminosity 
evolution if there has been no change in galaxy disk sizes.  

We have shown that we 1) are in good agreement with the observed 
magnitude -- size relation with time and the evolution in surface 
brightness, but 2) a number of the simulated disks are clearly undergoing 
a large change in size (Figs.~\ref{dmagsize} and \ref{evolsize}).  Our 
ability to follow individual galaxies with time, unlike the observations, 
allows us to interpret the evolution in the magnitude -- size plane.  We 
find that the evolution is dependent on mass, with our massive galaxies 
undergoing larger changes in size than magnitude.  

Our dwarf galaxies undergo the least change in size, though there is 
no immediate theoretical expectation that this should be the case.  
Recent results \citep{fakhouri10} show that the rate of growth of 
halos has little dependence on mass, and thus low mass galaxies should 
grow just as much in size as their more massive counterparts.  We  
conclude that a larger sample is needed to study if low mass galaxies 
truly grow less than more massive galaxies.  On the other hand, the 
evolution in magnitude for dwarf disk galaxies is dependent on their 
bursty SF history, but an individual dwarf can undergo significant 
dimming of at least 2 magnitudes.  A dichotomy of evolution with 
mass has been seen in observations, with low mass galaxies exhibiting
significantly more dimming since $z$=1 than more massive galaxies
\citep[e.g.,][]{melbourne07,kanwar08}.  

The halo properties for these galaxies ($\lambda,c,m_d,j_d$) predict 
a scale length that is generally shorter than the scale lengths 
derived from the simulated light profiles.  Fig.~\ref{ml} demonstrates 
that the light and mass distributions yield scale lengths in good 
agreement, so the discrepancy between $R_d$ predicted for NFW halos 
and from our light profiles cannot be due to a discrepancy between 
light and mass profiles.  Rather, using $\lambda_g$, the spin of the 
gas in the halo, brings the predicted and measured scale lengths into 
decent agreement.  Thus, the angular momentum distribution of the 
gas from which the stars are forming has been modified from that of 
the dark matter \citep[see also][]{sharma05, chen03}.  \citet{roskar10} 
have demonstrated that accreted gas is torqued by the hot halo after 
entering the virial radius, potentially preventing counter-rotating 
and low angular momentum gas from reaching the disk.  A full 
examination of how this process occurs is beyond the scope of this 
paper and left for future work.

Having determined that these galaxies are growing in size with time, 
we can investigate this growth in stellar mass.  Like previous 
theoretical models \citep{brook06, somerville08, far09, dutton10b}, we find 
that these galaxies grow in such a way as to stay along approximately 
the same stellar mass -- size relation with time.  Hence, at a fixed 
stellar mass, there is little change back to $z$=1 in the size of 
galaxies.  

Due to the fact that our larger galaxies have grown since $z$=1, we do 
not have the ability to directly compare results at a fixed size.  
In particular, we do not have a population of simulated disks at $z$=1 
with scale length greater than 3 kpc, though these larger disks do exist 
observationally.  Presumably we might generate these larger disks if 
we simulate higher mass halos.  However, it is not clear from this study 
how these larger disks would evolve to $z$=0.  Observationally, the number 
density of disk galaxies shows no evidence for change back to $z$=1 
\citep{lilly98, rav04, sargent07, kanwar08}.  In conjunction with the 
apparent lack of evolution in the size function of disks, this fact has 
been interpreted to mean that either galaxies are done growing by $z$=1, 
or that the rate of destruction of galaxies at given size must be equally 
matched by the rate of growth of galaxies into that size.  The fact that 
our galaxies are growing, while still matching the size-magnitude evolution, 
suggests that galaxies are moving into and out of a given bin in size 
during this time interval.  It is beyond the scope of this work to answer 
what happens to these larger galaxies, though we may speculate.  There are 
two possibilities that arise due to mergers; perhaps the larger disks 
undergo mergers and become early type galaxies \citep{bell07}, or perhaps 
mergers cause them to become bulge dominated disks.  As the majority of 
size-magnitude evolution studies have focused on pure disks or disks with 
small bulges, this would remove them from the sample being considered 
\citep{sargent07}.

Finally, we demonstrate that the simulated galaxy disks follow SFR -- 
mass and sSFR -- mass relations similar to observations.  While the 
growth of dark matter halos is nearly self-similar with mass, observations 
demonstrate that some process must break this self-similarity in the baryons 
\citep[e.g.,][]{benson03, cw09, schaye10}.  In the mass range of galaxy 
halos presented here, SN feedback is the process that regulates SF
as a function of halo mass, and likely drives gas outflows that vary 
as a function of mass (though outflow rates are left for future work). 
The regulation of SF as a function of mass leads to the reproduction of 
the observed stellar mass -- metallicity relationship \citep{Brooks07}, 
and the SFR -- mass relation shown here that is responsible for the 
reproduction of the size -- luminosity relation back to $z$=1.  
A galaxy at a fixed M$_*$ at $z$=1 will have a higher SFR than 
a $z$=0 counterpart, and thus a higher luminosity as well.  In conjunction 
with the weak evolution of the stellar mass -- size relation, this tells 
us that the difference in luminosity observed at a fixed size between 
$z$=1 and $z$=0 is due to the fact that the $z$=1 galaxy has a higher 
SFR and luminosity.

\acknowledgments

We would like to thank S. Miller and R. Ellis for use of their $z$=1
disk scale length data.  
AB acknowledges support from the Sherman Fairchild Foundation.  
AS was supported by the Gordon \& Betty Moore Foundation.  
FG and TQ were funded by NSF AST-0908499.
FG acknowledges support from a
Theodore Dunham grant, HST GO-1125, NSF grant AST-0607819 and NASA
ATP NNX08AG84G.
PJ acknowledges support from the W. M. Keck Foundation.
Simulations were run at TACC, ARSC, and NAS.



\begin{thebibliography}{174}
\expandafter\ifx\csname natexlab\endcsname\relax\def\natexlab#1{#1}\fi

\bibitem[{{Abadi} {et~al.}(2003){Abadi}, {Navarro}, {Steinmetz}, \&
  {Eke}}]{Abadi03}
{Abadi}, M.~G., {Navarro}, J.~F., {Steinmetz}, M., \& {Eke}, V.~R. 2003, \apj,
  597, 21

\bibitem[{{Barden} {et~al.}(2005){Barden}, {Rix}, {Somerville}, {Bell},
  {H{\"a}u{\ss}ler}, {Peng}, {Borch}, {Beckwith}, {Caldwell}, {Heymans},
  {Jahnke}, {Jogee}, {McIntosh}, {Meisenheimer}, {S{\'a}nchez}, {Wisotzki}, \&
  {Wolf}}]{barden05}
{Barden}, M. {et~al.} 2005, \apj, 635, 959

\bibitem[{{Barnes} \& {Efstathiou}(1987)}]{barnes87}
{Barnes}, J., \& {Efstathiou}, G. 1987, \apj, 319, 575

\bibitem[{{Bell} {et~al.}(2007){Bell}, {Zheng}, {Papovich}, {Borch}, {Wolf}, \&
  {Meisenheimer}}]{bell07}
{Bell}, E.~F., {Zheng}, X.~Z., {Papovich}, C., {Borch}, A., {Wolf}, C., \&
  {Meisenheimer}, K. 2007, \apj, 663, 834

\bibitem[{{Benson} {et~al.}(2003){Benson}, {Bower}, {Frenk}, {Lacey}, {Baugh},
  \& {Cole}}]{benson03}
{Benson}, A.~J., {Bower}, R.~G., {Frenk}, C.~S., {Lacey}, C.~G., {Baugh},
  C.~M., \& {Cole}, S. 2003, \apj, 599, 38

\bibitem[{{Blanton} \& {Roweis}(2007)}]{kcorrect}
{Blanton}, M.~R., \& {Roweis}, S. 2007, \aj, 133, 734

\bibitem[{{Boissier} {et~al.}(2010){Boissier}, {Buat}, \&
  {Ilbert}}]{boissier10}
{Boissier}, S., {Buat}, V., \& {Ilbert}, O. 2010, ArXiv e-prints

\bibitem[{{Booth} {et~al.}(2007){Booth}, {Theuns}, \& {Okamoto}}]{booth07}
{Booth}, C.~M., {Theuns}, T., \& {Okamoto}, T. 2007, \mnras, 376, 1588

\bibitem[{{Boselli} {et~al.}(2001){Boselli}, {Gavazzi}, {Donas}, \&
  {Scodeggio}}]{boselli01}
{Boselli}, A., {Gavazzi}, G., {Donas}, J., \& {Scodeggio}, M. 2001, \aj, 121,
  753

\bibitem[{{Bouche} {et~al.}(2009){Bouche}, {Dekel}, {Genzel}, {Genel},
  {Cresci}, {Forster Schreiber}, {Shapiro}, {Davies}, \& {Tacconi}}]{bouche10}
{Bouche}, N. {et~al.} 2009, ArXiv e-prints

\bibitem[{{Bouwens} \& {Silk}(2002)}]{bs02}
{Bouwens}, R., \& {Silk}, J. 2002, \apj, 568, 522

\bibitem[{{Brinchmann} {et~al.}(2004){Brinchmann}, {Charlot}, {White},
  {Tremonti}, {Kauffmann}, {Heckman}, \& {Brinkmann}}]{brinchmann04}
{Brinchmann}, J., {Charlot}, S., {White}, S.~D.~M., {Tremonti}, C.,
  {Kauffmann}, G., {Heckman}, T., \& {Brinkmann}, J. 2004, \mnras, 351, 1151

\bibitem[{{Brook} {et~al.}(2006){Brook}, {Kawata}, {Martel}, {Gibson}, \&
  {Bailin}}]{brook06}
{Brook}, C.~B., {Kawata}, D., {Martel}, H., {Gibson}, B.~K., \& {Bailin}, J.
  2006, \apj, 639, 126

\bibitem[{{Brooks}(2010)}]{bashfest}
{Brooks}, A. 2010, ArXiv e-prints, 1003.3882

\bibitem[{{Brooks} {et~al.}(2007){Brooks}, {Governato}, {Booth}, {Willman},
  {Gardner}, {Wadsley}, {Stinson}, \& {Quinn}}]{Brooks07}
{Brooks}, A.~M., {Governato}, F., {Booth}, C.~M., {Willman}, B., {Gardner},
  J.~P., {Wadsley}, J., {Stinson}, G., \& {Quinn}, T. 2007, \apjl, 655, L17

\bibitem[{{Brooks} {et~al.}(2009){Brooks}, {Governato}, {Quinn}, {Brook}, \&
  {Wadsley}}]{brooks09}
{Brooks}, A.~M., {Governato}, F., {Quinn}, T., {Brook}, C.~B., \& {Wadsley}, J.
  2009, \apj, 694, 396

\bibitem[{{Bullock} {et~al.}(2001){Bullock}, {Dekel}, {Kolatt}, {Kravtsov},
  {Klypin}, {Porciani}, \& {Primack}}]{bullock01}
{Bullock}, J.~S., {Dekel}, A., {Kolatt}, T.~S., {Kravtsov}, A.~V., {Klypin},
  A.~A., {Porciani}, C., \& {Primack}, J.~R. 2001, \apj, 555, 240

\bibitem[{{Calzetti}(2001)}]{calzetti01}
{Calzetti}, D. 2001, \pasp, 113, 1449

\bibitem[{{Ceverino} \& {Klypin}(2009)}]{ceverino09}
{Ceverino}, D., \& {Klypin}, A. 2009, \apj, 695, 292

\bibitem[{{Chen} {et~al.}(2003){Chen}, {Jing}, \& {Yoshikaw}}]{chen03}
{Chen}, D.~N., {Jing}, Y.~P., \& {Yoshikaw}, K. 2003, \apj, 597, 35

\bibitem[{{Conroy} \& {Wechsler}(2009)}]{cw09}
{Conroy}, C., \& {Wechsler}, R.~H. 2009, \apj, 696, 620

\bibitem[{{Courteau} {et~al.}(2007){Courteau}, {Dutton}, {van den Bosch},
  {MacArthur}, {Dekel}, {McIntosh}, \& {Dale}}]{courteau07}
{Courteau}, S., {Dutton}, A.~A., {van den Bosch}, F.~C., {MacArthur}, L.~A.,
  {Dekel}, A., {McIntosh}, D.~H., \& {Dale}, D.~A. 2007, \apj, 671, 203

\bibitem[{{Cowie} \& {Barger}(2008)}]{cb08}
{Cowie}, L.~L., \& {Barger}, A.~J. 2008, \apj, 686, 72

\bibitem[{{Daddi} {et~al.}(2007){Daddi}, {Dickinson}, {Morrison}, {Chary},
  {Cimatti}, {Elbaz}, {Frayer}, {Renzini}, {Pope}, {Alexander}, {Bauer},
  {Giavalisco}, {Huynh}, {Kurk}, \& {Mignoli}}]{Daddi07}
{Daddi}, E. {et~al.} 2007, \apj, 670, 156

\bibitem[{{Damen} {et~al.}(2009{\natexlab{a}}){Damen}, {F{\"o}rster Schreiber},
  {Franx}, {Labb{\'e}}, {Toft}, {van Dokkum}, \& {Wuyts}}]{damen09b}
{Damen}, M., {F{\"o}rster Schreiber}, N.~M., {Franx}, M., {Labb{\'e}}, I.,
  {Toft}, S., {van Dokkum}, P.~G., \& {Wuyts}, S. 2009{\natexlab{a}}, \apj,
  705, 617

\bibitem[{{Damen} {et~al.}(2009{\natexlab{b}}){Damen}, {Labb{\'e}}, {Franx},
  {van Dokkum}, {Taylor}, \& {Gawiser}}]{damen09a}
{Damen}, M., {Labb{\'e}}, I., {Franx}, M., {van Dokkum}, P.~G., {Taylor},
  E.~N., \& {Gawiser}, E.~J. 2009{\natexlab{b}}, \apj, 690, 937

\bibitem[{{de Blok} {et~al.}(2008){de Blok}, {Walter}, {Brinks},
  {Trachternach}, {Oh}, \& {Kennicutt}}]{thingsrot}
{de Blok}, W.~J.~G., {Walter}, F., {Brinks}, E., {Trachternach}, C., {Oh}, S.,
  \& {Kennicutt}, R.~C. 2008, \aj, 136, 2648

\bibitem[{{de Jong}(1996{\natexlab{a}})}]{dejong96a}
{de Jong}, R.~S. 1996{\natexlab{a}}, \aap, 313, 45

\bibitem[{{de Jong}(1996{\natexlab{b}})}]{dejong96b}
---. 1996{\natexlab{b}}, \aap, 313, 377

\bibitem[{{Dekel} \& {Silk}(1986)}]{DS86}
{Dekel}, A., \& {Silk}, J. 1986, \apj, 303, 39

\bibitem[{{D'Onghia} \& {Burkert}(2004)}]{donghia04}
{D'Onghia}, E., \& {Burkert}, A. 2004, \apjl, 612, L13

\bibitem[{{D'Onghia} {et~al.}(2006){D'Onghia}, {Burkert}, {Murante}, \&
  {Khochfar}}]{donghia06}
{D'Onghia}, E., {Burkert}, A., {Murante}, G., \& {Khochfar}, S. 2006, \mnras,
  372, 1525

\bibitem[{{D'Onghia} \& {Navarro}(2007)}]{dn07}
{D'Onghia}, E., \& {Navarro}, J.~F. 2007, \mnras, 380, L58

\bibitem[{{Driver} {et~al.}(2008){Driver}, {Popescu}, {Tuffs}, {Graham},
  {Liske}, \& {Baldry}}]{driver08}
{Driver}, S.~P., {Popescu}, C.~C., {Tuffs}, R.~J., {Graham}, A.~W., {Liske},
  J., \& {Baldry}, I. 2008, \apjl, 678, L101

\bibitem[{{Dunne} {et~al.}(2009){Dunne}, {Ivison}, {Maddox}, {Cirasuolo},
  {Mortier}, {Foucaud}, {Ibar}, {Almaini}, {Simpson}, \& {McLure}}]{dunne09}
{Dunne}, L. {et~al.} 2009, \mnras, 394, 3

\bibitem[{{Dutton}(2009)}]{dutton09}
{Dutton}, A.~A. 2009, \mnras, 396, 121

\bibitem[{{Dutton} {et~al.}(2010{\natexlab{a}}){Dutton}, {van den Bosch}, \&
  {Dekel}}]{dutton10a}
{Dutton}, A.~A., {van den Bosch}, F.~C., \& {Dekel}, A. 2010{\natexlab{a}},
  \mnras, 405, 1690

\bibitem[{{Dutton} {et~al.}(2010{\natexlab{b}}){Dutton}, {van den Bosch},
  {Faber}, {Simard}, {Kassin}, {Koo}, {Bundy}, {Huang}, {Weiner}, {Cooper},
  {Newman}, {Mozena}, \& {Koekemoer}}]{dutton10b}
{Dutton}, A.~A. {et~al.} 2010{\natexlab{b}}, ArXiv e-prints

\bibitem[{{Efstathiou} {et~al.}(1982){Efstathiou}, {Lake}, \&
  {Negroponte}}]{stability}
{Efstathiou}, G., {Lake}, G., \& {Negroponte}, J. 1982, \mnras, 199, 1069

\bibitem[{{Eke} {et~al.}(2001){Eke}, {Navarro}, \& {Steinmetz}}]{eke01}
{Eke}, V.~R., {Navarro}, J.~F., \& {Steinmetz}, M. 2001, \apj, 554, 114

\bibitem[{{Elbaz} {et~al.}(2007){Elbaz}, {Daddi}, {Le Borgne}, {Dickinson},
  {Alexander}, {Chary}, {Starck}, {Brandt}, {Kitzbichler}, {MacDonald},
  {Nonino}, {Popesso}, {Stern}, \& {Vanzella}}]{elbaz07}
{Elbaz}, D. {et~al.} 2007, \aap, 468, 33

\bibitem[{{Elmegreen} {et~al.}(2005){Elmegreen}, {Elmegreen}, {Vollbach},
  {Foster}, \& {Ferguson}}]{elmegreen05}
{Elmegreen}, B.~G., {Elmegreen}, D.~M., {Vollbach}, D.~R., {Foster}, E.~R., \&
  {Ferguson}, T.~E. 2005, \apj, 634, 101

\bibitem[{{Elmegreen} {et~al.}(2007){Elmegreen}, {Elmegreen}, {Ravindranath},
  \& {Coe}}]{elmegreen07}
{Elmegreen}, D.~M., {Elmegreen}, B.~G., {Ravindranath}, S., \& {Coe}, D.~A.
  2007, \apj, 658, 763

\bibitem[{{Erb} {et~al.}(2006){Erb}, {Shapley}, {Pettini}, {Steidel}, {Reddy},
  \& {Adelberger}}]{Erb06}
{Erb}, D.~K., {Shapley}, A.~E., {Pettini}, M., {Steidel}, C.~C., {Reddy},
  N.~A., \& {Adelberger}, K.~L. 2006, \apj, 644, 813

\bibitem[{{Fakhouri} {et~al.}(2010){Fakhouri}, {Ma}, \&
  {Boylan-Kolchin}}]{fakhouri10}
{Fakhouri}, O., {Ma}, C., \& {Boylan-Kolchin}, M. 2010, \mnras, 857

\bibitem[{{Fall} \& {Efstathiou}(1980)}]{FE80}
{Fall}, S.~M., \& {Efstathiou}, G. 1980, \mnras, 193, 189

\bibitem[{{Fathi} {et~al.}(2010){Fathi}, {Allen}, {Boch}, {Hatziminaoglou}, \&
  {Peletier}}]{fathi10}
{Fathi}, K., {Allen}, M., {Boch}, T., {Hatziminaoglou}, E., \& {Peletier},
  R.~F. 2010, \mnras, 822

\bibitem[{{Ferguson} {et~al.}(2004){Ferguson}, {Dickinson}, {Giavalisco},
  {Kretchmer}, {Ravindranath}, {Idzi}, {Taylor}, {Conselice}, {Fall},
  {Gardner}, {Livio}, {Madau}, {Moustakas}, {Papovich}, {Somerville},
  {Spinrad}, \& {Stern}}]{ferguson04}
{Ferguson}, H.~C. {et~al.} 2004, \apjl, 600, L107

\bibitem[{{Feulner} {et~al.}(2005){Feulner}, {Goranova}, {Drory}, {Hopp}, \&
  {Bender}}]{feulner05}
{Feulner}, G., {Goranova}, Y., {Drory}, N., {Hopp}, U., \& {Bender}, R. 2005,
  \mnras, 358, L1

\bibitem[{{Firmani} \& {Avila-Reese}(2009)}]{far09}
{Firmani}, C., \& {Avila-Reese}, V. 2009, \mnras, 396, 1675

\bibitem[{{F{\"o}rster Schreiber} {et~al.}(2006){F{\"o}rster Schreiber},
  {Genzel}, {Lehnert}, {Bouch{\'e}}, {Verma}, {Erb}, {Shapley}, {Steidel},
  {Davies}, {Lutz}, {Nesvadba}, {Tacconi}, {Eisenhauer}, {Abuter}, {Gilbert},
  {Gillessen}, \& {Sternberg}}]{fs06}
{F{\"o}rster Schreiber}, N.~M. {et~al.} 2006, \apj, 645, 1062

\bibitem[{{Gardner}(2001)}]{gardner01}
{Gardner}, J.~P. 2001, \apj, 557, 616

\bibitem[{{Gavazzi} \& {Scodeggio}(1996)}]{gavazzi96}
{Gavazzi}, G., \& {Scodeggio}, M. 1996, \aap, 312, L29

\bibitem[{{Geha} {et~al.}(2006){Geha}, {Blanton}, {Masjedi}, \&
  {West}}]{Geha06}
{Geha}, M., {Blanton}, M.~R., {Masjedi}, M., \& {West}, A.~A. 2006,
  astro-ph/0608295

\bibitem[{{Giavalisco} {et~al.}(2004){Giavalisco}, {Ferguson}, {Koekemoer},
  {Dickinson}, {Alexander}, {Bauer}, {Bergeron}, {Biagetti}, {Brandt},
  {Casertano}, {Cesarsky}, {Chatzichristou}, {Conselice}, {Cristiani}, {Da
  Costa}, {Dahlen}, {de Mello}, {Eisenhardt}, {Erben}, {Fall}, {Fassnacht},
  {Fosbury}, {Fruchter}, {Gardner}, {Grogin}, {Hook}, {Hornschemeier}, {Idzi},
  {Jogee}, {Kretchmer}, {Laidler}, {Lee}, {Livio}, {Lucas}, {Madau},
  {Mobasher}, {Moustakas}, {Nonino}, {Padovani}, {Papovich}, {Park},
  {Ravindranath}, {Renzini}, {Richardson}, {Riess}, {Rosati}, {Schirmer},
  {Schreier}, {Somerville}, {Spinrad}, {Stern}, {Stiavelli}, {Strolger},
  {Urry}, {Vandame}, {Williams}, \& {Wolf}}]{goods}
{Giavalisco}, M. {et~al.} 2004, \apjl, 600, L93

\bibitem[{{Gnedin} {et~al.}(2007){Gnedin}, {Weinberg}, {Pizagno}, {Prada}, \&
  {Rix}}]{gnedin07}
{Gnedin}, O.~Y., {Weinberg}, D.~H., {Pizagno}, J., {Prada}, F., \& {Rix}, H.
  2007, \apj, 671, 1115

\bibitem[{{Governato} {et~al.}(2010){Governato}, {Brook}, {Mayer}, {Brooks},
  {Rhee}, {Wadsley}, {Jonsson}, {Willman}, {Stinson}, {Quinn}, \&
  {Madau}}]{bulgeless}
{Governato}, F. {et~al.} 2010, \nat, 463, 203

\bibitem[{{Governato} {et~al.}(2009){Governato}, {Brook}, {Brooks}, {Mayer},
  {Willman}, {Jonsson}, {Stilp}, {Pope}, {Christensen}, {Wadsley}, \&
  {Quinn}}]{merger}
---. 2009, \mnras, 398, 312

\bibitem[{{Governato} {et~al.}(2008){Governato}, {Mayer}, \& {Brook}}]{g08}
{Governato}, F., {Mayer}, L., \& {Brook}, C. 2008, ArXiv e-prints, 801

\bibitem[{{Governato} {et~al.}(2004){Governato}, {Mayer}, {Wadsley}, {Gardner},
  {Willman}, {Hayashi}, {Quinn}, {Stadel}, \& {Lake}}]{governato04}
{Governato}, F. {et~al.} 2004, \apj, 607, 688

\bibitem[{{Governato} {et~al.}(2007){Governato}, {Willman}, {Mayer}, {Brooks},
  {Stinson}, {Valenzuela}, {Wadsley}, \& {Quinn}}]{G07}
{Governato}, F., {Willman}, B., {Mayer}, L., {Brooks}, A., {Stinson}, G.,
  {Valenzuela}, O., {Wadsley}, J., \& {Quinn}, T. 2007, \mnras, 374, 1479

\bibitem[{{Graham} \& {Worley}(2008)}]{graham08}
{Graham}, A.~W., \& {Worley}, C.~C. 2008, \mnras, 388, 1708

\bibitem[{{Guo} {et~al.}(2010){Guo}, {White}, {Li}, \&
  {Boylan-Kolchin}}]{guo09}
{Guo}, Q., {White}, S., {Li}, C., \& {Boylan-Kolchin}, M. 2010, \mnras, 404,
  1111

\bibitem[{{Guo} \& {White}(2008)}]{guo08}
{Guo}, Q., \& {White}, S.~D.~M. 2008, \mnras, 384, 2

\bibitem[{{Haardt} \& {Madau}(1996)}]{HM96}
{Haardt}, F., \& {Madau}, P. 1996, \apj, 461, 20

\bibitem[{{Harfst} {et~al.}(2006){Harfst}, {Theis}, \& {Hensler}}]{harfst06}
{Harfst}, S., {Theis}, C., \& {Hensler}, G. 2006, \aap, 449, 509

\bibitem[{{Hetznecker} \& {Burkert}(2006)}]{hetznecker06}
{Hetznecker}, H., \& {Burkert}, A. 2006, \mnras, 370, 1905

\bibitem[{{Hopkins} {et~al.}(2009){Hopkins}, {Cox}, {Younger}, \&
  {Hernquist}}]{hopkinsbulgea}
{Hopkins}, P.~F., {Cox}, T.~J., {Younger}, J.~D., \& {Hernquist}, L. 2009,
  \apj, 691, 1168

\bibitem[{{Hultman} \& {Pharasyn}(1999)}]{hultman99}
{Hultman}, J., \& {Pharasyn}, A. 1999, \aap, 347, 769

\bibitem[{{Jones} {et~al.}(2009){Jones}, {Swinbank}, {Ellis}, {Richard}, \&
  {Stark}}]{jones09}
{Jones}, T., {Swinbank}, M., {Ellis}, R., {Richard}, J., \& {Stark}, D. 2009,
  ArXiv e-prints

\bibitem[{{Jonsson}(2006)}]{sunrise}
{Jonsson}, P. 2006, \mnras, 372, 2

\bibitem[{{Jonsson} {et~al.}(2010){Jonsson}, {Groves}, \& {Cox}}]{sunrise2}
{Jonsson}, P., {Groves}, B.~A., \& {Cox}, T.~J. 2010, \mnras, 403, 17

\bibitem[{{Kanwar} {et~al.}(2008){Kanwar}, {Simard}, {Schade}, \&
  {Gwyn}}]{kanwar08}
{Kanwar}, A., {Simard}, L., {Schade}, D., \& {Gwyn}, S.~D.~J. 2008, \apj, 682,
  907

\bibitem[{{Katz}(1992)}]{katz92}
{Katz}, N. 1992, \apj, 391, 502

\bibitem[{{Katz} {et~al.}(1994){Katz}, {Quinn}, {Bertschinger}, \&
  {Gelb}}]{katz94}
{Katz}, N., {Quinn}, T., {Bertschinger}, E., \& {Gelb}, J.~M. 1994, \mnras,
  270, L71+

\bibitem[{{Katz} \& {White}(1993)}]{KW93}
{Katz}, N., \& {White}, S.~D.~M. 1993, \apj, 412, 455

\bibitem[{{Kauffmann} {et~al.}(2003){Kauffmann}, {Heckman}, {White}, {Charlot},
  {Tremonti}, {Peng}, {Seibert}, {Brinkmann}, {Nichol}, {SubbaRao}, \&
  {York}}]{K03}
{Kauffmann}, G. {et~al.} 2003, \mnras, 341, 54

\bibitem[{{Kaufmann} {et~al.}(2007){Kaufmann}, {Mayer}, {Wadsley}, {Stadel}, \&
  {Moore}}]{kaufmann07}
{Kaufmann}, T., {Mayer}, L., {Wadsley}, J., {Stadel}, J., \& {Moore}, B. 2007,
  \mnras, 375, 53

\bibitem[{{Kennicutt} {et~al.}(2003){Kennicutt}, {Armus}, {Bendo}, {Calzetti},
  {Dale}, {Draine}, {Engelbracht}, {Gordon}, {Grauer}, {Helou}, {Hollenbach},
  {Jarrett}, {Kewley}, {Leitherer}, {Li}, {Malhotra}, {Regan}, {Rieke},
  {Rieke}, {Roussel}, {Smith}, {Thornley}, \& {Walter}}]{sings}
{Kennicutt}, Jr., R.~C. {et~al.} 2003, \pasp, 115, 928

\bibitem[{{Kere{\v s}} {et~al.}(2009){Kere{\v s}}, {Katz}, {Fardal},
  {Dav{\'e}}, \& {Weinberg}}]{keres09}
{Kere{\v s}}, D., {Katz}, N., {Fardal}, M., {Dav{\'e}}, R., \& {Weinberg},
  D.~H. 2009, \mnras, 395, 160

\bibitem[{{Komatsu} {et~al.}(2008){Komatsu}, {Dunkley}, {Nolta}, {Bennett},
  {Gold}, {Hinshaw}, {Jarosik}, {Larson}, {Limon}, {Page}, {Spergel},
  {Halpern}, {Hill}, {Kogut}, {Meyer}, {Tucker}, {Weiland}, {Wollack}, \&
  {Wright}}]{wmap5}
{Komatsu}, E. {et~al.} 2008, ArXiv e-prints, 803

\bibitem[{{Komatsu} {et~al.}(2010){Komatsu}, {Smith}, {Dunkley}, {Bennett},
  {Gold}, {Hinshaw}, {Jarosik}, {Larson}, {Nolta}, {Page}, {Spergel},
  {Halpern}, {Hill}, {Kogut}, {Limon}, {Meyer}, {Odegard}, {Tucker}, {Weiland},
  {Wollack}, \& {Wright}}]{wmap7}
---. 2010, ArXiv e-prints

\bibitem[{{Kroupa} {et~al.}(1993){Kroupa}, {Tout}, \& {Gilmore}}]{Kroupa}
{Kroupa}, P., {Tout}, C.~A., \& {Gilmore}, G. 1993, \mnras, 262, 545

\bibitem[{{Labb{\'e}} {et~al.}(2003){Labb{\'e}}, {Rudnick}, {Franx}, {Daddi},
  {van Dokkum}, {F{\"o}rster Schreiber}, {Kuijken}, {Moorwood}, {Rix},
  {R{\"o}ttgering}, {Trujillo}, {van der Wel}, {van der Werf}, \& {van
  Starkenburg}}]{labbe03}
{Labb{\'e}}, I. {et~al.} 2003, \apjl, 591, L95

\bibitem[{{Lara-L{\'o}pez} {et~al.}(2010){Lara-L{\'o}pez}, {Cepa},
  {Bongiovanni}, {P{\'e}rez Garc{\'{\i}}a}, {Ederoclite}, {Casta{\~n}eda},
  {Fern{\'a}ndez Lorenzo}, {P{\'o}vic}, \& {S{\'a}nchez-Portal}}]{laralopez10}
{Lara-L{\'o}pez}, M.~A. {et~al.} 2010, ArXiv e-prints

\bibitem[{{Leitherer} {et~al.}(1999){Leitherer}, {Schaerer}, {Goldader},
  {Gonz{\'a}lez Delgado}, {Robert}, {Kune}, {de Mello}, {Devost}, \&
  {Heckman}}]{starburst99a}
{Leitherer}, C. {et~al.} 1999, \apjs, 123, 3

\bibitem[{{Lilly} {et~al.}(1998){Lilly}, {Schade}, {Ellis}, {Le Fevre},
  {Brinchmann}, {Tresse}, {Abraham}, {Hammer}, {Crampton}, {Colless},
  {Glazebrook}, {Mallen-Ornelas}, \& {Broadhurst}}]{lilly98}
{Lilly}, S. {et~al.} 1998, \apj, 500, 75

\bibitem[{{MacArthur} {et~al.}(2003){MacArthur}, {Courteau}, \&
  {Holtzman}}]{macarthur03}
{MacArthur}, L.~A., {Courteau}, S., \& {Holtzman}, J.~A. 2003, \apj, 582, 689

\bibitem[{{MacArthur} {et~al.}(2008){MacArthur}, {Ellis}, {Treu}, {U}, {Bundy},
  \& {Moran}}]{macarthur08}
{MacArthur}, L.~A., {Ellis}, R.~S., {Treu}, T., {U}, V., {Bundy}, K., \&
  {Moran}, S. 2008, \apj, 680, 70

\bibitem[{{Maiolino} {et~al.}(2008){Maiolino}, {Nagao}, {Grazian}, {Cocchia},
  {Marconi}, {Mannucci}, {Cimatti}, {Pipino}, {Ballero}, {Calura}, {Chiappini},
  {Fontana}, {Granato}, {Matteucci}, {Pastorini}, {Pentericci}, {Risaliti},
  {Salvati}, \& {Silva}}]{maiolino08}
{Maiolino}, R. {et~al.} 2008, \aap, 488, 463

\bibitem[{{Maller} \& {Dekel}(2002)}]{maller02b}
{Maller}, A.~H., \& {Dekel}, A. 2002, \mnras, 335, 487

\bibitem[{{Maller} {et~al.}(2002){Maller}, {Dekel}, \&
  {Somerville}}]{maller02a}
{Maller}, A.~H., {Dekel}, A., \& {Somerville}, R. 2002, \mnras, 329, 423

\bibitem[{{Mannucci} {et~al.}(2010){Mannucci}, {Cresci}, {Maiolino}, {Marconi},
  \& {Gnerucci}}]{mannucci10}
{Mannucci}, F., {Cresci}, G., {Maiolino}, R., {Marconi}, A., \& {Gnerucci}, A.
  2010, ArXiv e-prints

\bibitem[{{Mao} {et~al.}(1998){Mao}, {Mo}, \& {White}}]{mao98}
{Mao}, S., {Mo}, H.~J., \& {White}, S.~D.~M. 1998, \mnras, 297, L71

\bibitem[{{Marri} \& {White}(2003)}]{marri03}
{Marri}, S., \& {White}, S.~D.~M. 2003, \mnras, 345, 561

\bibitem[{{Mayer} {et~al.}(2008){Mayer}, {Governato}, \& {Kaufmann}}]{mayer08}
{Mayer}, L., {Governato}, F., \& {Kaufmann}, T. 2008, ArXiv e-prints, 801

\bibitem[{{McKee} \& {Ostriker}(1977)}]{MO77}
{McKee}, C.~F., \& {Ostriker}, J.~P. 1977, \apj, 218, 148

\bibitem[{{Melbourne} {et~al.}(2007){Melbourne}, {Phillips}, {Harker}, {Novak},
  {Koo}, \& {Faber}}]{melbourne07}
{Melbourne}, J., {Phillips}, A.~C., {Harker}, J., {Novak}, G., {Koo}, D.~C., \&
  {Faber}, S.~M. 2007, \apj, 660, 81

\bibitem[{{Mo} {et~al.}(1998){Mo}, {Mao}, \& {White}}]{mmw}
{Mo}, H.~J., {Mao}, S., \& {White}, S.~D.~M. 1998, \mnras, 295, 319

\bibitem[{{Moster} {et~al.}(2010{\natexlab{a}}){Moster}, {Macci{\`o}},
  {Somerville}, {Johansson}, \& {Naab}}]{moster10b}
{Moster}, B.~P., {Macci{\`o}}, A.~V., {Somerville}, R.~S., {Johansson}, P.~H.,
  \& {Naab}, T. 2010{\natexlab{a}}, \mnras, 403, 1009

\bibitem[{{Moster} {et~al.}(2010{\natexlab{b}}){Moster}, {Somerville},
  {Maulbetsch}, {van den Bosch}, {Macci{\`o}}, {Naab}, \& {Oser}}]{moster10}
{Moster}, B.~P., {Somerville}, R.~S., {Maulbetsch}, C., {van den Bosch}, F.~C.,
  {Macci{\`o}}, A.~V., {Naab}, T., \& {Oser}, L. 2010{\natexlab{b}}, \apj, 710,
  903

\bibitem[{{Naab} {et~al.}(2007){Naab}, {Johansson}, {Ostriker}, \&
  {Efstathiou}}]{naab07}
{Naab}, T., {Johansson}, P.~H., {Ostriker}, J.~P., \& {Efstathiou}, G. 2007,
  \apj, 658, 710

\bibitem[{{Navarro} \& {Benz}(1991)}]{nb91}
{Navarro}, J.~F., \& {Benz}, W. 1991, \apj, 380, 320

\bibitem[{{Navarro} \& {Steinmetz}(2000)}]{ns00}
{Navarro}, J.~F., \& {Steinmetz}, M. 2000, \apj, 538, 477

\bibitem[{{Navarro} \& {White}(1994)}]{nw94}
{Navarro}, J.~F., \& {White}, S.~D.~M. 1994, \mnras, 267, 401

\bibitem[{{Noeske} {et~al.}(2007{\natexlab{a}}){Noeske}, {Faber}, {Weiner},
  {Koo}, {Primack}, {Dekel}, {Papovich}, {Conselice}, {Le Floc'h}, {Rieke},
  {Coil}, {Lotz}, {Somerville}, \& {Bundy}}]{noeske07b}
{Noeske}, K.~G. {et~al.} 2007{\natexlab{a}}, \apjl, 660, L47

\bibitem[{{Noeske} {et~al.}(2007{\natexlab{b}}){Noeske}, {Weiner}, {Faber},
  {Papovich}, {Koo}, {Somerville}, {Bundy}, {Conselice}, {Newman},
  {Schiminovich}, {Le Floc'h}, {Coil}, {Rieke}, {Lotz}, {Primack}, {Barmby},
  {Cooper}, {Davis}, {Ellis}, {Fazio}, {Guhathakurta}, {Huang}, {Kassin},
  {Martin}, {Phillips}, {Rich}, {Small}, {Willmer}, \& {Wilson}}]{noeske07a}
---. 2007{\natexlab{b}}, \apjl, 660, L43

\bibitem[{{Nordstr{\"o}m} {et~al.}(2004){Nordstr{\"o}m}, {Mayor}, {Andersen},
  {Holmberg}, {Pont}, {J{\o}rgensen}, {Olsen}, {Udry}, \&
  {Mowlavi}}]{nordstrom04}
{Nordstr{\"o}m}, B. {et~al.} 2004, \aap, 418, 989

\bibitem[{{Okamoto} {et~al.}(2005){Okamoto}, {Eke}, {Frenk}, \&
  {Jenkins}}]{okamoto05}
{Okamoto}, T., {Eke}, V.~R., {Frenk}, C.~S., \& {Jenkins}, A. 2005, \mnras,
  363, 1299

\bibitem[{{Oliver} {et~al.}(2010){Oliver}, {Frost}, {Farrah},
  {Gonzalez-Solares}, {Shupe}, {Henriques}, {Roseboom}, {Alfonso-Luis},
  {Babbedge}, {Frayer}, {Lencz}, {Lonsdale}, {Masci}, {Padgett}, {Polletta},
  {Rowan-Robinson}, {Siana}, {Smith}, {Surace}, \& {Vaccari}}]{oliver10}
{Oliver}, S. {et~al.} 2010, \mnras, 588

\bibitem[{{Pannella} {et~al.}(2009){Pannella}, {Carilli}, {Daddi}, {McCracken},
  {Owen}, {Renzini}, {Strazzullo}, {Civano}, {Koekemoer}, {Schinnerer},
  {Scoville}, {Smol{\v c}i{\'c}}, {Taniguchi}, {Aussel}, {Kneib}, {Ilbert},
  {Mellier}, {Salvato}, {Thompson}, \& {Willott}}]{pannella09}
{Pannella}, M. {et~al.} 2009, \apjl, 698, L116

\bibitem[{{Peebles}(1969)}]{peebles69}
{Peebles}, P.~J.~E. 1969, \apj, 155, 393

\bibitem[{{Peirani} {et~al.}(2004){Peirani}, {Mohayaee}, \& {de Freitas
  Pacheco}}]{peirani04}
{Peirani}, S., {Mohayaee}, R., \& {de Freitas Pacheco}, J.~A. 2004, \mnras,
  348, 921

\bibitem[{{Peng} {et~al.}(2002){Peng}, {Ho}, {Impey}, \& {Rix}}]{galfit}
{Peng}, C.~Y., {Ho}, L.~C., {Impey}, C.~D., \& {Rix}, H. 2002, \aj, 124, 266

\bibitem[{{Piontek} \& {Steinmetz}(2009)}]{piontek09a}
{Piontek}, F., \& {Steinmetz}, M. 2009, ArXiv e-prints

\bibitem[{{Pizagno} {et~al.}(2005){Pizagno}, {Prada}, {Weinberg}, {Rix},
  {Harbeck}, {Grebel}, {Bell}, {Brinkmann}, {Holtzman}, \& {West}}]{pizagno05}
{Pizagno}, J. {et~al.} 2005, \apj, 633, 844

\bibitem[{{Pontzen} {et~al.}(2010){Pontzen}, {Deason}, {Governato}, {Pettini},
  {Wadsley}, {Quinn}, {Brooks}, {Bellovary}, \& {Fynbo}}]{pontzen10}
{Pontzen}, A. {et~al.} 2010, \mnras, 402, 1523

\bibitem[{{Pontzen} {et~al.}(2008){Pontzen}, {Governato}, {Pettini}, {Booth},
  {Stinson}, {Wadsley}, {Brooks}, {Quinn}, \& {Haehnelt}}]{pontzen08}
---. 2008, \mnras, 390, 1349

\bibitem[{{Ravindranath} {et~al.}(2004){Ravindranath}, {Ferguson}, {Conselice},
  {Giavalisco}, {Dickinson}, {Chatzichristou}, {de Mello}, {Fall}, {Gardner},
  {Grogin}, {Hornschemeier}, {Jogee}, {Koekemoer}, {Kretchmer}, {Livio},
  {Mobasher}, \& {Somerville}}]{rav04}
{Ravindranath}, S. {et~al.} 2004, \apjl, 604, L9

\bibitem[{{Reshetnikov} {et~al.}(2003){Reshetnikov}, {Dettmar}, \&
  {Combes}}]{reshetnikov03}
{Reshetnikov}, V.~P., {Dettmar}, R., \& {Combes}, F. 2003, \aap, 399, 879

\bibitem[{{Rix} {et~al.}(2004){Rix}, {Barden}, {Beckwith}, {Bell}, {Borch},
  {Caldwell}, {H{\"a}ussler}, {Jahnke}, {Jogee}, {McIntosh}, {Meisenheimer},
  {Peng}, {Sanchez}, {Somerville}, {Wisotzki}, \& {Wolf}}]{gems}
{Rix}, H. {et~al.} 2004, \apjs, 152, 163

\bibitem[{{Robertson} {et~al.}(2004){Robertson}, {Yoshida}, {Springel}, \&
  {Hernquist}}]{Robertson04}
{Robertson}, B., {Yoshida}, N., {Springel}, V., \& {Hernquist}, L. 2004, \apj,
  606, 32

\bibitem[{{Robertson} \& {Kravtsov}(2008)}]{robertson08}
{Robertson}, B.~E., \& {Kravtsov}, A.~V. 2008, \apj, 680, 1083

\bibitem[{{Roche} {et~al.}(1998){Roche}, {Ratnatunga}, {Griffiths}, {Im}, \&
  {Naim}}]{roche98}
{Roche}, N., {Ratnatunga}, K., {Griffiths}, R.~E., {Im}, M., \& {Naim}, A.
  1998, \mnras, 293, 157

\bibitem[{{Rodighiero} {et~al.}(2010){Rodighiero}, {Cimatti}, {Gruppioni},
  {Popesso}, {Andreani}, {Altieri}, {Aussel}, {Berta}, {Bongiovanni},
  {Brisbin}, {Cava}, {Cepa}, {Daddi}, {Dominguez-Sanchez}, {Elbaz}, {Fontana},
  {Forster Schreiber}, {Franceschini}, {Genzel}, {Grazian}, {Lutz}, {Magdis},
  {Magliocchetti}, {Magnelli}, {Maiolino}, {Mancini}, {Nordon}, {Perez Garcia},
  {Poglitsch}, {Santini}, {Sanchez-Portal}, {Pozzi}, {Riguccini}, {Saintonge},
  {Shao}, {Sturm}, {Tacconi}, {Valtchanov}, {Wetzstein}, \&
  {Wieprecht}}]{rodighiero10}
{Rodighiero}, G. {et~al.} 2010, ArXiv e-prints

\bibitem[{{Ro{\v s}kar} {et~al.}(2010){Ro{\v s}kar}, {Debattista}, {Brooks},
  {Quinn}, {Brook}, {Governato}, {Dalcanton}, \& {Wadsley}}]{roskar10}
{Ro{\v s}kar}, R., {Debattista}, V.~P., {Brooks}, A.~M., {Quinn}, T.~R.,
  {Brook}, C.~B., {Governato}, F., {Dalcanton}, J.~J., \& {Wadsley}, J. 2010,
  \mnras, 408, 783

\bibitem[{{Saitoh} {et~al.}(2008){Saitoh}, {Daisaka}, {Kokubo}, {Makino},
  {Okamoto}, {Tomisaka}, {Wada}, \& {Yoshida}}]{saitoh08}
{Saitoh}, T.~R., {Daisaka}, H., {Kokubo}, E., {Makino}, J., {Okamoto}, T.,
  {Tomisaka}, K., {Wada}, K., \& {Yoshida}, N. 2008, \pasj, 60, 667

\bibitem[{{Sales} {et~al.}(2010){Sales}, {Navarro}, {Schaye}, {Dalla Vecchia},
  {Springel}, \& {Booth}}]{sales10}
{Sales}, L.~V., {Navarro}, J.~F., {Schaye}, J., {Dalla Vecchia}, C.,
  {Springel}, V., \& {Booth}, C.~M. 2010, ArXiv e-prints

\bibitem[{{Sales} {et~al.}(2009){Sales}, {Navarro}, {Schaye}, {Dalla Vecchia},
  {Springel}, {Haas}, \& {Helmi}}]{sales09}
{Sales}, L.~V., {Navarro}, J.~F., {Schaye}, J., {Dalla Vecchia}, C.,
  {Springel}, V., {Haas}, M.~R., \& {Helmi}, A. 2009, \mnras, 399, L64

\bibitem[{{Salim} {et~al.}(2007){Salim}, {Rich}, {Charlot}, {Brinchmann},
  {Johnson}, {Schiminovich}, {Seibert}, {Mallery}, {Heckman}, {Forster},
  {Friedman}, {Martin}, {Morrissey}, {Neff}, {Small}, {Wyder}, {Bianchi},
  {Donas}, {Lee}, {Madore}, {Milliard}, {Szalay}, {Welsh}, \& {Yi}}]{salim07}
{Salim}, S. {et~al.} 2007, \apjs, 173, 267

\bibitem[{{S{\'a}nchez-Janssen} {et~al.}(2010){S{\'a}nchez-Janssen},
  {M{\'e}ndez-Abreu}, \& {Aguerri}}]{sanchezjanssen10}
{S{\'a}nchez-Janssen}, R., {M{\'e}ndez-Abreu}, J., \& {Aguerri}, J.~A.~L. 2010,
  \mnras, 406, L65

\bibitem[{{Sargent} {et~al.}(2007){Sargent}, {Carollo}, {Lilly}, {Scarlata},
  {Feldmann}, {Kampczyk}, {Koekemoer}, {Scoville}, {Kneib}, {Leauthaud},
  {Massey}, {Rhodes}, {Tasca}, {Capak}, {McCracken}, {Porciani}, {Renzini},
  {Taniguchi}, {Thompson}, \& {Sheth}}]{sargent07}
{Sargent}, M.~T. {et~al.} 2007, \apjs, 172, 434

\bibitem[{{Scannapieco} {et~al.}(2010){Scannapieco}, {Gadotti}, {Jonsson}, \&
  {White}}]{scannapieco10}
{Scannapieco}, C., {Gadotti}, D.~A., {Jonsson}, P., \& {White}, S.~D.~M. 2010,
  \mnras, L99+

\bibitem[{{Scannapieco} {et~al.}(2006){Scannapieco}, {Tissera}, {White}, \&
  {Springel}}]{cecelia0}
{Scannapieco}, C., {Tissera}, P.~B., {White}, S.~D.~M., \& {Springel}, V. 2006,
  \mnras, 371, 1125

\bibitem[{{Scannapieco} {et~al.}(2008){Scannapieco}, {Tissera}, {White}, \&
  {Springel}}]{cecilia1}
---. 2008, \mnras, 389, 1137

\bibitem[{{Scannapieco} {et~al.}(2009){Scannapieco}, {White}, {Springel}, \&
  {Tissera}}]{cecilia2}
{Scannapieco}, C., {White}, S.~D.~M., {Springel}, V., \& {Tissera}, P.~B. 2009,
  \mnras, 396, 696

\bibitem[{{Schade} {et~al.}(1996){Schade}, {Carlberg}, {Yee}, {Lopez-Cruz}, \&
  {Ellingson}}]{schade96}
{Schade}, D., {Carlberg}, R.~G., {Yee}, H.~K.~C., {Lopez-Cruz}, O., \&
  {Ellingson}, E. 1996, \apjl, 465, L103+

\bibitem[{{Schaye} {et~al.}(2010){Schaye}, {Dalla Vecchia}, {Booth}, {Wiersma},
  {Theuns}, {Haas}, {Bertone}, {Duffy}, {McCarthy}, \& {van de
  Voort}}]{schaye10}
{Schaye}, J. {et~al.} 2010, \mnras, 402, 1536

\bibitem[{{Schiminovich} {et~al.}(2007){Schiminovich}, {Wyder}, {Martin},
  {Johnson}, {Salim}, {Seibert}, {Treyer}, {Budav{\'a}ri}, {Hoopes},
  {Zamojski}, {Barlow}, {Forster}, {Friedman}, {Morrissey}, {Neff}, {Small},
  {Bianchi}, {Donas}, {Heckman}, {Lee}, {Madore}, {Milliard}, {Rich}, {Szalay},
  {Welsh}, \& {Yi}}]{schiminovich07}
{Schiminovich}, D. {et~al.} 2007, \apjs, 173, 315

\bibitem[{{Schombert}(2006)}]{schombert06}
{Schombert}, J.~M. 2006, \aj, 131, 296

\bibitem[{{Scoville} {et~al.}(2007){Scoville}, {Aussel}, {Brusa}, {Capak},
  {Carollo}, {Elvis}, {Giavalisco}, {Guzzo}, {Hasinger}, {Impey}, {Kneib},
  {LeFevre}, {Lilly}, {Mobasher}, {Renzini}, {Rich}, {Sanders}, {Schinnerer},
  {Schminovich}, {Shopbell}, {Taniguchi}, \& {Tyson}}]{cosmos}
{Scoville}, N. {et~al.} 2007, \apjs, 172, 1

\bibitem[{{Shapiro} {et~al.}(2008){Shapiro}, {Genzel}, {F{\"o}rster Schreiber},
  {Tacconi}, {Bouch{\'e}}, {Cresci}, {Davies}, {Eisenhauer}, {Johansson},
  {Krajnovi{\'c}}, {Lutz}, {Naab}, {Arimoto}, {Arribas}, {Cimatti}, {Colina},
  {Daddi}, {Daigle}, {Erb}, {Hernandez}, {Kong}, {Mignoli}, {Onodera},
  {Renzini}, {Shapley}, \& {Steidel}}]{shapiro08}
{Shapiro}, K.~L. {et~al.} 2008, \apj, 682, 231

\bibitem[{{Sharma} \& {Steinmetz}(2005)}]{sharma05}
{Sharma}, S., \& {Steinmetz}, M. 2005, \apj, 628, 21

\bibitem[{{Shen} {et~al.}(2003){Shen}, {Mo}, {White}, {Blanton}, {Kauffmann},
  {Voges}, {Brinkmann}, \& {Csabai}}]{shen03}
{Shen}, S., {Mo}, H.~J., {White}, S.~D.~M., {Blanton}, M.~R., {Kauffmann}, G.,
  {Voges}, W., {Brinkmann}, J., \& {Csabai}, I. 2003, \mnras, 343, 978

\bibitem[{{Simard} {et~al.}(1999){Simard}, {Koo}, {Faber}, {Sarajedini},
  {Vogt}, {Phillips}, {Gebhardt}, {Illingworth}, \& {Wu}}]{simard99}
{Simard}, L. {et~al.} 1999, \apj, 519, 563

\bibitem[{{Somerville} {et~al.}(2008){Somerville}, {Barden}, {Rix}, {Bell},
  {Beckwith}, {Borch}, {Caldwell}, {H{\"a}u{\ss}ler}, {Heymans}, {Jahnke},
  {Jogee}, {McIntosh}, {Meisenheimer}, {Peng}, {S{\'a}nchez}, {Wisotzki}, \&
  {Wolf}}]{somerville08}
{Somerville}, R.~S. {et~al.} 2008, \apj, 672, 776

\bibitem[{{Sommer-Larsen} {et~al.}(1999){Sommer-Larsen}, {Gelato}, \&
  {Vedel}}]{sommerlarsen99}
{Sommer-Larsen}, J., {Gelato}, S., \& {Vedel}, H. 1999, \apj, 519, 501

\bibitem[{{Sommer-Larsen} {et~al.}(2003){Sommer-Larsen}, {G{\"o}tz}, \&
  {Portinari}}]{sommerlarsen03}
{Sommer-Larsen}, J., {G{\"o}tz}, M., \& {Portinari}, L. 2003, \apj, 596, 47

\bibitem[{{Springel} \& {Hernquist}(2003)}]{springel03}
{Springel}, V., \& {Hernquist}, L. 2003, \mnras, 339, 289

\bibitem[{{Stadel}(2001)}]{pkdgrav}
{Stadel}, J.~G. 2001, PhD thesis, AA(UNIVERSITY OF WASHINGTON)

\bibitem[{{Stark} {et~al.}(2008){Stark}, {Swinbank}, {Ellis}, {Dye}, {Smail},
  \& {Richard}}]{stark08}
{Stark}, D.~P., {Swinbank}, A.~M., {Ellis}, R.~S., {Dye}, S., {Smail}, I.~R.,
  \& {Richard}, J. 2008, \nat, 455, 775

\bibitem[{{Steinmetz} \& {Navarro}(1999)}]{sn99}
{Steinmetz}, M., \& {Navarro}, J.~F. 1999, \apj, 513, 555

\bibitem[{{Steinmetz} \& {White}(1997)}]{sw97}
{Steinmetz}, M., \& {White}, S.~D.~M. 1997, \mnras, 288, 545

\bibitem[{{Stinson} {et~al.}(2006){Stinson}, {Seth}, {Katz}, {Wadsley},
  {Governato}, \& {Quinn}}]{Stinson06}
{Stinson}, G., {Seth}, A., {Katz}, N., {Wadsley}, J., {Governato}, F., \&
  {Quinn}, T. 2006, \mnras, 373, 1074

\bibitem[{{Tasker} \& {Bryan}(2008)}]{tasker08}
{Tasker}, E.~J., \& {Bryan}, G.~L. 2008, \apj, 673, 810

\bibitem[{{Thacker} \& {Couchman}(2000)}]{thacker00b}
{Thacker}, R.~J., \& {Couchman}, H.~M.~P. 2000, \apj, 545, 728

\bibitem[{{Thacker} \& {Couchman}(2001)}]{thacker01}
---. 2001, \apjl, 555, L17

\bibitem[{{Trujillo} \& {Aguerri}(2004)}]{trujillo04}
{Trujillo}, I., \& {Aguerri}, J.~A.~L. 2004, \mnras, 355, 82

\bibitem[{{Trujillo} {et~al.}(2006){Trujillo}, {F{\"o}rster Schreiber},
  {Rudnick}, {Barden}, {Franx}, {Rix}, {Caldwell}, {McIntosh}, {Toft},
  {H{\"a}ussler}, {Zirm}, {van Dokkum}, {Labb{\'e}}, {Moorwood},
  {R{\"o}ttgering}, {van der Wel}, {van der Werf}, \& {van
  Starkenburg}}]{trujillo06}
{Trujillo}, I. {et~al.} 2006, \apj, 650, 18

\bibitem[{{Trujillo} \& {Pohlen}(2005)}]{trujillo05}
{Trujillo}, I., \& {Pohlen}, M. 2005, \apjl, 630, L17

\bibitem[{{van den Bosch}(1998)}]{vdb98}
{van den Bosch}, F.~C. 1998, \apj, 507, 601

\bibitem[{{van den Bosch} {et~al.}(2002){van den Bosch}, {Abel}, {Croft},
  {Hernquist}, \& {White}}]{vdb02}
{van den Bosch}, F.~C., {Abel}, T., {Croft}, R.~A.~C., {Hernquist}, L., \&
  {White}, S.~D.~M. 2002, \apj, 576, 21

\bibitem[{{van den Bosch} {et~al.}(2003){van den Bosch}, {Abel}, \&
  {Hernquist}}]{vdb03}
{van den Bosch}, F.~C., {Abel}, T., \& {Hernquist}, L. 2003, \mnras, 346, 177

\bibitem[{{van den Bosch} {et~al.}(2001){van den Bosch}, {Burkert}, \&
  {Swaters}}]{vdb01}
{van den Bosch}, F.~C., {Burkert}, A., \& {Swaters}, R.~A. 2001, \mnras, 326,
  1205

\bibitem[{{van Zee}(2000)}]{vanzee00}
{van Zee}, L. 2000, \aj, 119, 2757

\bibitem[{{V{\'a}zquez} \& {Leitherer}(2005)}]{starburst99b}
{V{\'a}zquez}, G.~A., \& {Leitherer}, C. 2005, \apj, 621, 695

\bibitem[{{Vitvitska} {et~al.}(2002){Vitvitska}, {Klypin}, {Kravtsov},
  {Wechsler}, {Primack}, \& {Bullock}}]{vitvitska}
{Vitvitska}, M., {Klypin}, A.~A., {Kravtsov}, A.~V., {Wechsler}, R.~H.,
  {Primack}, J.~R., \& {Bullock}, J.~S. 2002, \apj, 581, 799

\bibitem[{{Vogt} {et~al.}(1996){Vogt}, {Forbes}, {Phillips}, {Gronwall},
  {Faber}, {Illingworth}, \& {Koo}}]{vogt96}
{Vogt}, N.~P., {Forbes}, D.~A., {Phillips}, A.~C., {Gronwall}, C., {Faber},
  S.~M., {Illingworth}, G.~D., \& {Koo}, D.~C. 1996, \apjl, 465, L15+

\bibitem[{{Wadsley} {et~al.}(2004){Wadsley}, {Stadel}, \& {Quinn}}]{gasoline}
{Wadsley}, J.~W., {Stadel}, J., \& {Quinn}, T. 2004, New Astronomy, 9, 137

\bibitem[{{White}(1984)}]{white84}
{White}, S.~D.~M. 1984, \apj, 286, 38

\bibitem[{{White} \& {Rees}(1978)}]{wr78}
{White}, S.~D.~M., \& {Rees}, M.~J. 1978, \mnras, 183, 341

\bibitem[{{Wright} {et~al.}(2009){Wright}, {Larkin}, {Law}, {Steidel},
  {Shapley}, \& {Erb}}]{wright09}
{Wright}, S.~A., {Larkin}, J.~E., {Law}, D.~R., {Steidel}, C.~C., {Shapley},
  A.~E., \& {Erb}, D.~K. 2009, \apj, 699, 421

\bibitem[{{York} {et~al.}(2000){York}, {Adelman}, {Anderson}, {Anderson},
  {Annis}, {Bahcall}, {Bakken}, {Barkhouser}, {Bastian}, {Berman}, {Boroski},
  {Bracker}, {Briegel}, {Briggs}, {Brinkmann}, {Brunner}, {Burles}, {Carey},
  {Carr}, {Castander}, {Chen}, {Colestock}, {Connolly}, {Crocker}, {Csabai},
  {Czarapata}, {Davis}, {Doi}, {Dombeck}, {Eisenstein}, {Ellman}, {Elms},
  {Evans}, {Fan}, {Federwitz}, {Fiscelli}, {Friedman}, {Frieman}, {Fukugita},
  {Gillespie}, {Gunn}, {Gurbani}, {de Haas}, {Haldeman}, {Harris}, {Hayes},
  {Heckman}, {Hennessy}, {Hindsley}, {Holm}, {Holmgren}, {Huang}, {Hull},
  {Husby}, {Ichikawa}, {Ichikawa}, {Ivezi{\'c}}, {Kent}, {Kim}, {Kinney},
  {Klaene}, {Kleinman}, {Kleinman}, {Knapp}, {Korienek}, {Kron}, {Kunszt},
  {Lamb}, {Lee}, {Leger}, {Limmongkol}, {Lindenmeyer}, {Long}, {Loomis},
  {Loveday}, {Lucinio}, {Lupton}, {MacKinnon}, {Mannery}, {Mantsch}, {Margon},
  {McGehee}, {McKay}, {Meiksin}, {Merelli}, {Monet}, {Munn}, {Narayanan},
  {Nash}, {Neilsen}, {Neswold}, {Newberg}, {Nichol}, {Nicinski}, {Nonino},
  {Okada}, {Okamura}, {Ostriker}, {Owen}, {Pauls}, {Peoples}, {Peterson},
  {Petravick}, {Pier}, {Pope}, {Pordes}, {Prosapio}, {Rechenmacher}, {Quinn},
  {Richards}, {Richmond}, {Rivetta}, {Rockosi}, {Ruthmansdorfer}, {Sandford},
  {Schlegel}, {Schneider}, {Sekiguchi}, {Sergey}, {Shimasaku}, {Siegmund},
  {Smee}, {Smith}, {Snedden}, {Stone}, {Stoughton}, {Strauss}, {Stubbs},
  {SubbaRao}, {Szalay}, {Szapudi}, {Szokoly}, {Thakar}, {Tremonti}, {Tucker},
  {Uomoto}, {Vanden Berk}, {Vogeley}, {Waddell}, {Wang}, {Watanabe},
  {Weinberg}, {Yanny}, \& {Yasuda}}]{sdss}
{York}, D.~G. {et~al.} 2000, \aj, 120, 1579

\bibitem[{{Zavala} {et~al.}(2008){Zavala}, {Okamoto}, \& {Frenk}}]{zavala08}
{Zavala}, J., {Okamoto}, T., \& {Frenk}, C.~S. 2008, \mnras, 387, 364

\end{thebibliography}
\end{document}